\documentclass{article}
\usepackage[T1]{fontenc}
\usepackage[utf8]{inputenc}
\usepackage{ismir}
\usepackage{amsmath,cite,url}
\usepackage{graphicx}
\usepackage{color}
\usepackage{booktabs}
\usepackage{comment}
\usepackage{siunitx}

\title{Assessing Factual Music Comprehension \\ in Large Audio Language Models}

\threeauthors
  {Daniel Chenyu Lin} {Cornell University \\ \texttt{cl2765@cornell.edu}}
  {Michael Freeman} {Cornell University \\ \texttt{mf867@cornell.edu}}
  {John Thickstun} {Cornell University \\ \texttt{jthickstun@cornell.edu}}

\def\authorname{D. Lin, M. Freeman, and J. Thickstun}

\begin{document}

\maketitle

\begin{abstract}
Large audio language models (LALMs) leverage multimodal representations to generate open-ended answers to natural language queries about audio. In this paper, we (1) provide empirical evidence that assessment of LALMs using the popular MusicQA dataset fails to measure whether a model's responses about music are factually correct, and (2) develop a new protocol for assessing the music comprehension capabilities of LALMs. Specifically, we propose an evaluation protocol that prompts a LALM for factually verifiable information, and parses its open-ended response into a structured format that can be objectively assessed using Precision, Recall, and F1 scores. Using this protocol, we define a benchmark consisting of six factual information retrieval tasks defined on three diverse datasets: MusicNet, the Free Music Archive, and OverClocked ReMix. We benchmark nine recent LALMs, including frontier models like Gemini and the latest open models like Music Flamingo, and release the suite of evaluation scripts at \url{https://github.com/DCL2004/LALM-Eval} to facilitate benchmarking of new LALMs.
\end{abstract}

\section{Introduction}
\label{sec:introduction}

\begin{figure*}[t!]
    \centering
    \includegraphics[width=1\linewidth]{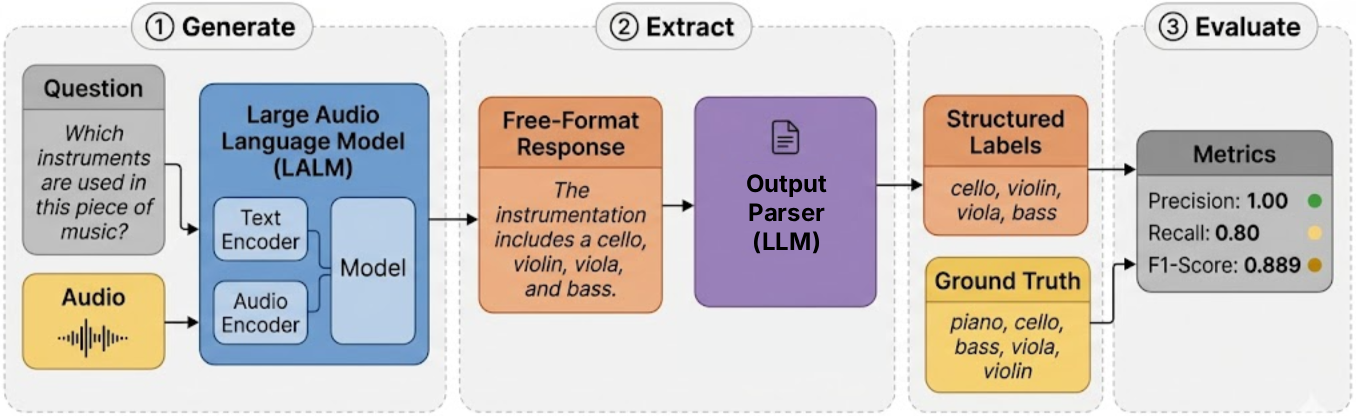}
    \caption{
    Our factuality framework for converting a labeled dataset into a benchmark for LALMs. A LALM first predicts open-ended text in response to a prompt for factual information. A large language model then performs output parsing under strict rules to canonicalize this free-form response into structured labels. These extracted labels are compared to ground-truth labels to compute factuality metrics such as precision, recall, and F1-score, enabling direct, interpretable evaluation of factual correctness.}
    \label{fig:figure_1}
    \vspace{-1mm}
\end{figure*}

Large Audio Language Models (LALMs) are an emerging family of multimodal generative models that consume both text and audio inputs and produce text output. LALMs are trained to answer natural language questions about audio recordings, including music, making these models a promising general-purpose tool for music information retrieval tasks such as captioning, tagging, and interactive question answering. In this work, we seek to understand how well recent LALMs perform on a variety of classic music information retrieval tasks, and develop a reproducible benchmark for measuring progress in the ongoing development of this family of models.

The standard protocol for measuring a LALM's music comprehension capabilities is to pose questions to the model from a musical question-answering dataset such as MusicQA~\cite{mu-llama}, and compare the model's free-form text answers with the dataset's reference answers using text similarity metrics such as BERTScore \cite{bertscore}. There is growing evidence that these datasets and metrics are inadequate for assessing the music comprehension capabilities of modern LALMs~\cite{llark, captioning,muchomusic, areyoureallylistening}. In Section~\ref{sec:qa} we corroborate these findings, showing that text similarity scores for LALM generated answers to MusicQA questions do not materially change when the audio input to the LALM is replaced with a random different audio clip from the MusicQA dataset.

In Section~\ref{sec:factuality}, we develop a protocol for creating targeted assessments of specific factual music comprehension capabilities in LALMs. This protocol prompts LALMs with factual questions about music, such as naming the instruments, or identifying a musical genre. We then use a language model to convert unstructured LALM outputs into a structured format that can be compared with ground-truth annotations using Precision, Recall, and F1 scores. In contrast to free-form QA, we show that this factual evaluation protocol effectively measures audio comprehension capabilities in LALMs. Because each prompt targets a specific fact, the protocol provides a fine-grained and interpretable assessment of LALM performance across various music information retrieval tasks.

We evaluate factual music comprehension performance of nine recent LALMs across six factual information retrieval tasks. We use the MusicNet dataset to assess instrument recognition and composer identification \cite{musicnet}. We use the Free Music Archive (FMA) to assess genre classification \cite{fma}. Additionally, we identify OverClocked ReMix as an additional open dataset that can be used to measure regional style, mood, and time signature identification.

Our contributions are as follows:
\begin{itemize}
    \item We demonstrate that the free-form captioning protocol fails to measure music comprehension in a study of eight LALMs using the MusicQA dataset.
    
    \item We propose a factual evaluation protocol that converts free-form LALM responses into structured labels. Labels are directly compared to ground-truth annotations using Precision, Recall, and F1 scores.
    
    \item Using this evaluation protocol, we benchmark the performance of nine recent music LALMs across six information retrieval tasks on three datasets.
\end{itemize}

Together, these findings identify a new approach for evaluating LALMs and lay a foundation for further benchmarking of  developments in this family of models.

\section{Related Work}
\label{related_work}

Early work on LALMs for music borrowed evaluation methods from the Natural Language Processing (NLP) literature~\cite{mu-llama,llark,mumu-llama}, comparing LALM generated answers against reference responses on the captioning subset of MusicQA using BLEU\cite{bleu}, BLEU-4, METEOR~\cite{meteor}, ROUGE, and BERTScore~\cite{bertscore}; noting concerns about these metrics, LLark and the Audio Flamingo series~\cite{audio-flamingo,audio-flamingo2,audio-flamingo3} additionally report CIDEr~\cite{cider}.
All of these metrics are reference-based n-gram or embedding similarity scores; as we show in Section~\ref{sec:qa}, none of them distinguish a model conditioned on the correct audio input from audio conditioned on random audio input. 

A separate line of work evaluates LALMs through multiple-choice question answering (MCQA), whereby the model is prompted with a question together with candidate answers and asked to choose an answer from the list~\cite{muchomusic,mmau}. Results are scored based on accuracy; Audio Flamingo 3 and Music Flamingo~\cite{music-flamingo} both report MCQA accuracy. This format conflates audio comprehension with linguistic test-taking skills. MCQA accuracy scores are sensitive to choice order and option phrasing\cite{robustness}. Indeed, language models without any audio input are shown to answer up to 56.4\% of MuChoMusic questions correctly, comparable or better than current LALMs \cite{areyoureallylistening}.

The RUListening protocol patches MCQA by using a Perceptual Index computed from text-only LLM probabilities to identify questions that already require audio, and synthesizing hard distractor answers for the remaining questions~\cite{areyoureallylistening}. Our work shares the diagnosis that LALM evaluations can fail to isolate audio-grounded behavior, but our remedy is different. Rather than prompt the LALM with multiple-choice responses that are susceptible to linguistic heuristics, we ask open-ended questions and rely on a powerful LLM to format the LALM's responses.

Audio Flamingo 3 and Music Flamingo\cite{music-flamingo} adopt an LLM-as-a-judge protocol of \cite{lm-judge}, prompting GPT-4o or GPT-5 to assign a 1--10 quality score to each candidate compared with a reference. This protocol raises reproducibility concerns specific to closed-source judges and still rewards proximity to a single reference caption even though a given recording may admit many equally valid descriptions emphasizing different facets of the music.

\section{Free-Form Question Answering}
\label{sec:qa}

Question-answering datasets such as MusicQA probe the capability of LALMs to respond to open-ended questions about audio. MusicQA  poses five content-specific questions about each audio clip in the dataset, as well as four generic questions asked about every audio clip: ``Describe the music,'' ``Describe the music in detail,'' ``What do you hear in the audio,'' and ``What can be inferred from the audio.'' The model responds in natural language to these queries and its free-form responses are compared using text-similarity metrics to reference responses from the MusicQA dataset. In this section, we present a series of experiments showing that comparing free-form~QA responses to reference text using the MusicQA dataset has no discriminative power for evaluating the capabilities of LALMs.

\subsection{Baseline}

To contextualize the free-form QA performance of LALMs, our baseline experiment pairs each question with a random, unrelated music recording from the same dataset. This measures how models perform when they receive irrelevant audio; intuitively, all LALMs should outperform this baseline when they are given the correct audio. This baseline is analogous to the procedure studied for MCQA in \cite{areyoureallylistening}, which replaces audio prompts with Gaussian noise. We opt to replace the audio with random audio from the same dataset to keep the input in-distribution, thereby avoiding pathological behaviors.

\begin{table*}[t!]
\caption{Aggregate similarity metrics between reference and response text on the MusicQA-Jamendo Free-Form QA dataset (5,040 QA pairs). NLP metrics barely distinguish between the quality of answers to queries using the correct song as input, versus a random song from the dataset. With the exception of CIDEr, all metrics are bound by $1.0$. CIDEr in this context multiplies embedding similarity by $10$, which bounds it within $0$ and $10$.}
\label{tab:musicqa-jamendo}
\begin{center}
\begin{tabular}{ll rrrrrr}
\toprule
\multicolumn{8}{l}{\textbf{MusicQA-Jamendo}} \\
\midrule
\textbf{Model} & \textbf{Prompt} & \textbf{BLEU} & \textbf{BLEU-4} & \textbf{METEOR} & \textbf{ROUGE} & \textbf{BERTScore} & \textbf{CIDEr} \\
\midrule
LTU-AS          & Correct     & 0.2487 & 0.1643 & 0.2723 & 0.3144 & 0.8847 & 0.4731 \\
                & Random      & 0.2505 & 0.1640 & 0.2749 & 0.3183 & 0.8863 & 0.4255 \\
\midrule
MU-LLaMA        & Correct     & 0.3015 & 0.2084 & 0.3891 & 0.4609 & 0.8997 & 0.3288 \\
                & Random      & 0.2906 & 0.1961 & 0.3779 & 0.4529 & 0.8968 & 0.2858 \\
\midrule
LLaMA-Adapter   & Correct     & 0.2001 & 0.1321 & 0.3270 & 0.5201 & 0.8915 & 0.1063 \\
                & Random      & 0.1951 & 0.1260 & 0.3163 & 0.5096 & 0.8889 & 0.1013 \\
\midrule
SALMONN         & Correct     & 0.2950 & 0.2197 & 0.3505 & 0.4184 & 0.8985 & 0.9262 \\
                & Random      & 0.2700 & 0.2041 & 0.3270 & 0.3836 & 0.8918 & 0.9232 \\
\midrule
Qwen2-Audio     & Correct     & 0.1511 & 0.0925 & 0.1651 & 0.1982 & 0.8502 & 0.2181 \\
                & Random      & 0.1423 & 0.0836 & 0.1550 & 0.1891 & 0.8484 & 0.2004 \\
\midrule
Audio-Flamingo 3 & Correct    & 0.1993 & 0.1010 & 0.1580 & 0.1682 & 0.8831 & 0.0915 \\
                & Random      & 0.1826 & 0.0927 & 0.1545 & 0.1597 & 0.8814 & 0.0778 \\
\midrule
Qwen3-Omni      & Correct     & 0.0537 & 0.0374 & 0.1559 & 0.5506 & 0.8712 & 0.0147 \\
                & Random      & 0.0500 & 0.0331 & 0.1443 & 0.5179 & 0.8644 & 0.0098 \\
\midrule
Music-Flamingo  & Correct     & 0.2062 & 0.1397 & 0.2232 & 0.2399 & 0.8799 & 0.4540 \\
                & Random      & 0.1816 & 0.1186 & 0.2000 & 0.2157 & 0.8752 & 0.3652 \\
\midrule
GPT-4.1-mini    & Paraphrase  & 0.5956 & 0.4648 & 0.5968 & 0.5663 & 0.9581 & 1.7632 \\
               & Adversarial & 0.7413 & 0.6614 & 0.7739 & 0.7609 & 0.9608 & 3.8270 \\
\bottomrule
\end{tabular}
\end{center}
\vspace{-3mm}
\end{table*}

\subsection{Skyline (Paraphrase)}

As a conceptual upper bound on performance, we also conduct a skyline experiment to emulate a near perfect response. We paraphrase the reference text using \texttt{ChatGPT-4.1-mini}. Paraphrased responses mimic the reference answers, but written differently. 
We assume that a modern LLM is capable of rephrasing text without changing its meaning; we verify this by asking musicians to inspect and validate a sampling of paraphrased outputs.

\subsection{Adversarial Rewrites}

We consider an adversarial experiment to examine how metrics behave when for a response that is deliberately incorrect. Instead of asking \texttt{ChatGPT-4.1-mini} to preserve the meaning of the reference text, we ask it to \textit{change} the meaning in as few edits as possible. This rewrite should therefore be lexically similar to the original, but carry different meaning. Changing the meaning renders the rewrite incorrect with respect to the original audio paired with it. A good evaluation metric should assign low scores to these adversarial answers. 

\subsection{Results}

We report the standard metrics used to compare LALM responses to reference text using the Music QA dataset in Table~\ref{tab:musicqa-jamendo}, comparing responses using ``Correct'' song inputs to responses when the model is provided with a ``Random'' unrelated song from the dataset. Ideally, we would expect models provided with correct song inputs to significantly outperform models provided with a random song. Strong models should converge towards the skyline paraphrased answers. Across all eight models and all six metrics, we find that this is not the case. In many cases, the random-audio condition scores \emph{higher} than the correct-audio condition.

CIDEr shows the most consistent gaps between these two experimental settings, but they remain small for every model, and far below the skyline paraphrased metric performance. All metrics are fooled by the adversarial rewrites, which remain lexically similar to the reference despite contradicting it factually.

We identify two possible reasons that text similarity metrics could fail to measure quality of LALM answers for MusicQA. One possibility is that the metrics are too sensitive to surface-level fluency of responses; the success of the adversarially rewritten answers suggests that this is part of the problem. An alternative explanation is that the questions posed by MusicQA are too vague, and the audio has insufficient information to meaningfully predict the reference answers. Likely both issues contribute to the negative results presented in Table~\ref{tab:musicqa-jamendo}. Regardless of the explanation, it is clear that MusicQA is not a viable benchmark for assessing LALMs in the music domain. This motivates our work in Section~\ref{sec:factuality} to develop an alternative protocol for benchmarking LALMs that can substantively assess music comprehension in these models.

\section{Factual Question Answering}
\label{sec:factuality}

In light of difficulties evaluating free-form QA responses, we propose a \emph{factual QA} evaluation protocol for probing LALMs on matters of fact, e.g., instrument recognition or composer identification. In principle, answers to these questions can be evaluated using straightforward accuracy metrics. In practice, LALMs produce free-form text responses that require further analysis to determine correctness. We find that many LALMs are incapable of following instructions to produce responses in a specific format. Therefore, we propose a protocol whereby (1) questions are posed to a LALM about specific factual labels on the content of audio, (2) the free-form outputs of the LALM are parsed using a strong LLM into a structured format, and (3) these structured responses are compared with ground-truth labels using simple metrics. The structure of this factuality protocol is shown in Figure~\ref{fig:figure_1}.

In the remainder of this section, we describe an implementation of factual QA using the MusicNet, FMA, and OverClocked ReMix datasets. Using this implementation, we benchmark the capabilities of nine modern LALMs. We follow the convention established by FEVER \cite{fever} and 
FActScore \cite{factscore}, considering ``factuality'' in the sense of verifiability against a designated reference, in this case, labels obtained from the MusicNet, FMA, and OverClocked ReMix datasets. Some labels (e.g., genre, mood) reflect sociocultural conventions rather than objective acoustic properties of the audio; we treat these as factual in the sense that they conform to an agreed-upon ground truth established by domain experts.

\subsection{Output Parsing}
\label{keyword_extraction}

We use an LLM to convert free-form text responses from a LALM into a canonical structured form that can be compared to ground-truth labels. Whereas the evaluation of free-form responses in Section~\ref{sec:qa} conflates factual content with stylistic cues, our goal here is to extract explicit labels in a structured, machine-checkable form. For example, if the model responds to a query about genre with the free-form text ``\emph{The genre of the song is rock},'' we would like to parse this response into a categorical label \emph{rock}; if the model responds to a query about instrument content with, ``\emph{Instruments: double bass and horns},'' we would like to parse this response into a set of labels \{\emph{bass, horn}\}.

The parser is governed by a strict rule set that restricts extraction to terms the model explicitly states and forbids inference beyond the response.
(full rules in Appendix~\ref{keyword_extraction_rules}) 
For tasks with unambiguous canonical vocabularies, including instrumentation, composer, genre, and time signature, the parser returns exact mentions, with light dataset-side normalization to mitigate surface-form mismatches (e.g., \emph{contrabass} and \emph{double bass} both map to \emph{bass}). For tasks whose labels admit genuine semantic overlap, such as mood and regional style, we instead present the parser with the full closed list of dataset labels and ask it to select the labels best reflected by the model's response (e.g., \emph{Egyptian} $\to$ \emph{Middle Eastern}); the same strict rules govern this variant. 

\begin{figure*}[t!]
    \centering
    \includegraphics[width=1\linewidth]{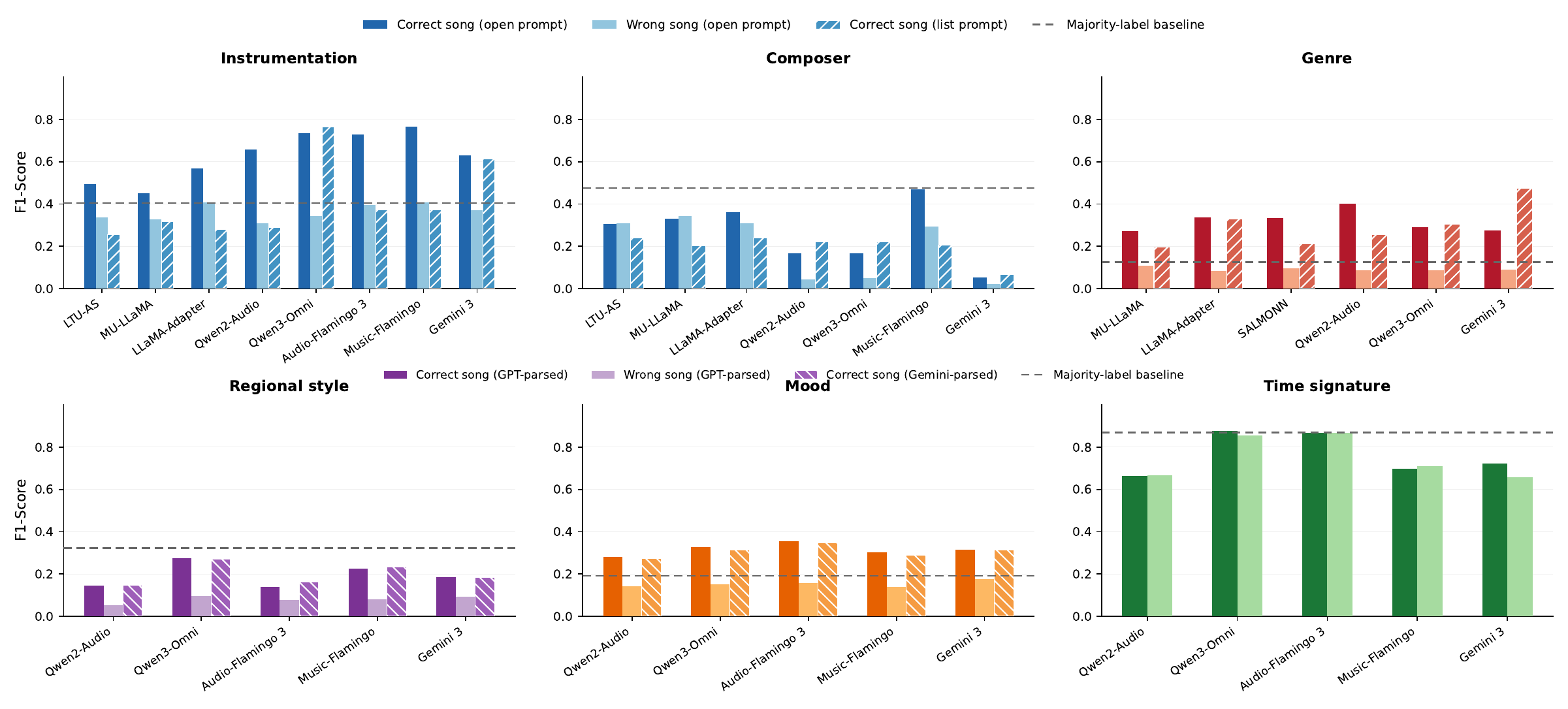}
    \caption{Factual QA F1-scores across six music understanding tasks. \textbf{Top row:} Instrumentation recognition and composer classification on MusicNet, and genre classification on FMA. Each model is evaluated under three conditions: correct audio with an open-ended prompt (dark), random audio with an open-ended prompt (light), and correct audio with a list prompt that enumerates all possible answers (hatched). \textbf{Bottom row:} Regional style, mood, and time signature classification on OverClocked ReMix. Regional and mood results include a third bar showing correct-audio F1 when using Gemini as the label parser instead of GPT (hatched), to assess parser sensitivity.
    Gray dashed lines mark the F1 of always predicting each dataset's most common label.
    For all tasks except time signature, models score substantially higher with correct audio than with random audio, confirming that factual metrics capture genuine music understanding. Time signature scores show minimal correct-vs-random separation, suggesting models default to predicting 4/4 regardless of input.}
    \label{fig:factuality_graphs}
\end{figure*}

To verify that results are not sensitive to the LLM used for parsing, we parse using \texttt{ChatGPT-4.1-mini} and \texttt{Gemini-3.1-flash-light} for the two most ambiguous tasks in our benchmark: regional style and mood classification. Results using these two different LLMs to parse outputs are in close agreement: see Figure~\ref{fig:factuality_graphs}. 
Detail of the prompts used to instruct an LLM to perform this parsing task is described in the Appendix \ref{keyword_extraction}.

\subsection{Evaluation Metrics for Parsed Labels}
\label{eval_metrics}
One complication of factual evaluation is that models may fail to respond to the question or equivocate, outputting zero or multiple answers to questions that have a single ground-truth label. A contributing factor to this behavior is architectural: some LALMs process audio in chunks, with the potential to produce a different answers for each chunk. Even unchunked models sometimes hedge their responses, declining to respond or giving several possible answers. To account for these behaviors, we adopt Precision, Recall, and F1 Score, which allow us to treat model predictions as sets of labels and measure both correctness and hedging in responses. In the special case where a model emits exactly one answer per question, F1 reduces to Accuracy. Meanwhile, unchunked models may also give ambiguous output within which multiple responses are given without specific preference. An example is given in Table ~\ref{tab:multiple_keywords_extraction}.

\begin{table}[t!]
\caption{An example for multiple labels parsing. In the table, MU-LLaMA gives two possible answers to the question about genre without preference. Here we accept all labels generated by the model and compare them to the ground truth using Precision/Recall/F1-score. In this example, Precision = 0.5, Recall = 1, F1-Score = 0.667.}
\label{tab:multiple_keywords_extraction}
\begin{center}
\small
\resizebox{\columnwidth}{!}{%
\begin{tabular}{lp{0.75\columnwidth}}
\toprule
\textbf{Source} & \textbf{Text} \\
\midrule
Question & What genre does this piece of music fall under? \\
Label & Pop \\
Output & This piece of music falls under the genre of pop/soft rock. \\
Parsed Labels & Pop, Rock \\
\bottomrule
\end{tabular}
}
\end{center}
\vspace{-3mm}
\end{table}

\section{Experimental Results}
\label{sec:factuality_results}

We use MusicNet, Free Music Archive (FMA), and a relatively unexplored dataset, OverClocked ReMix, to support a suite of experiments that implement the factuality protocol developed in Section~\ref{sec:factuality}.
Because MusicNet and FMA are well-established datasets, several of the evaluated models include them in their training data (e.g., SALMONN trains on MusicNet, and Audio Flamingo 3 and Music Flamingo train on FMA; see Appendix~\ref{model_details}). To avoid evaluating models on their own training data, we omit model–dataset pairs with known overlap. Additionally, because the older LALMs already underperform on the classic tasks, we evaluate only the newer models on OverClocked ReMix.
In Section~\ref{sec:musicnet} we evaluate instrument recognition and classical composer classification using the MusicNet dataset. This dataset consists of \num{330} full-length classical recordings, each annotated with a composer label (one of ten composers) and a list of instruments. 
In Section~\ref{sec:fma} we evaluate genre classification using the FMA-Small subset of FMA. This dataset consists of \num{8000} thirty-second clips evenly distributed across eight top-level genres: hip-hop, pop, folk, experimental, rock, international, electronic, and instrumental (\num{1000} clips per genre).

For experiments on MusicNet and FMA we consider two different strategies for prompting a LALM: an \emph{open prompt} and a \emph{list prompt}. In the open prompting strategy we ask the language model an open-ended question, e.g., ``What genre does this piece of music fall under?'' In the list prompt strategy, we provide the model with a full list of valid responses, e.g., ``Among hip-hop, pop, folk, experimental, rock, international, electronic, instrumental, what is the genre of this song?'' 
We consider the list prompt are comparable to MCQA prompting as we provide the model with possible labels yet trying to be more robust from model guessing only based on the question prompting as shown in \cite{areyoureallylistening}.
We initially hypothesized that the list strategy would be easier;
this appears to be the case for newer LALMs (e.g., Gemini) but older models are notably confused when prompted with a list of options. Music-Flamingo is one recent LALM that is surprisingly sensitive to the list prompt.
We discuss additional prompting strategies in the Appendix \ref{prompting_studies}.

Finally, we in Section~\ref{sec:ocr} we evaluate regional style and mood recognition as well as time signature identification using the open OverClocked ReMix dataset. This is a relatively unexplored dataset in the MIR community, consisting of a curated collection of \num{4270} high-quality video game remixes with extensive annotations by remix community experts. We make use of just a small subset of these annotation tags in the present work; the website associated with this data contains extensive text descriptions, reviews, and discussion boards attached to each audio recording that could prove fruitful for further training and evaluation of LALMs as these multimodal models become more sophisticated.\footnote{\url{https://ocremix.org/}} 

Figure 2 also plots the F1 of always predicting each dataset's most common label: piano, Beethoven, 4/4, World, and Energetic; FMA-Small is genre-balanced, so its baseline is chance (0.125).

\subsection{MusicNet}\label{sec:musicnet}

For instrument recognition, models tend to output many labels, resulting in high recall scores but low precision: most correct instruments are included, but predictions also contain false positives. This phenomenon is less pronounced for classification tasks but still degrades performance, as seen in both composer and genre classification experiments. For composer classification, we observe that music composed by Bach and Beethoven comprise a large fraction of the MusicNet dataset. Music Flamingo achieves the highest score in this task by intelligently guessing between a restricted subset of composers: Bach, Mozart, and Beethoven. LLaMA-Adapter is able to perform the second best by guessing Beethoven for the majority of the entries. In contrast, models that output more diverse responses such as Qwen3-Omni and Gemini 3 are less successful. No model exceeds the always-Beethoven baseline of 0.476.

\subsection{Free Music Archive}\label{sec:fma}

FMA provides a hierarchical taxonomy for genre classification, in which each top-level genre contains multiple subgenres organized in a tree structure; we only use top-level genre labels (see the Appendix \ref{same_node_fma} for an application of our framework to full genre trees in FMA). For top-level genre classification, there are eight possible genre labels distributed evenly over the FMA-Small dataset; a model that randomly guesses genre labels should score around $0.125$. Nevertheless, the baseline random audio experiment scores slightly lower than  $0.125$ because the models sometimes respond with no answer, or equivocate among multiple answers.

Like the MusicNet tasks, LALM performance tends to degrade when given a list prompt for genre classification, with the notable exception of Gemini 3 which is able to use the list of options to significantly assist its analysis.

\subsection{OverClocked ReMix}\label{sec:ocr}

We apply the factual QA protocol to subsets of OverClocked ReMix annotated with regional style and mood labels as described in Table~\ref{tab:ocr}. Because regional style and mood are imprecise labels, we instruct the parser to map LALM outputs to the closest valid label where possible, rather than requiring an exact match (e.g.\ \emph{Egyptian} $\to$ Middle Eastern; \emph{South African} $\to$ African).
This raises the concern that different parser LLMs may map the same response to different categories. For example, we could imagine that GPT might map the output \emph{upbeat} to \emph{energetic} whereas Gemini maps the same output to \emph{happy}.

To measure parser sensitivity, we re-ran the extraction step using both GPT-4.1-mini and Gemini~3 for both the regional style and mood identification tasks. The assessed performance using each parser is nearly identical for every model. The largest single discrepancy occurs for Audio-Flamingo~3 on regional style: this model emits many unrelated outputs that the parsers reduce to empty labels, and the resulting empty-label majority amplifies disagreement on the remaining answers. Even so, the absolute difference between the two parsers is quite small (see the hatched bars in Figure~\ref{fig:factuality_graphs}).

For regional style, we initially expected models to confuse \emph{South American} with \emph{Latin American} due to geographic overlap, but in practice every model distinguishes the two reliably. Confusions are model-specific: Gemini~3 tends to confuse Chinese with Japanese music, and Qwen3-Omni tends to confuse Indian with Middle Eastern music. Neither confusion appears in the other models. All models fall below the always-World baseline, but World is a catch-all that no LALM predicts; the correct-versus-random gap shows models do distinguish regional styles.

\begin{table}[t]
\centering
\caption{OverClocked ReMix Styles and Moods}
\label{tab:ocr}
\small

\begin{minipage}{0.48\columnwidth}
\centering
\begin{tabular}{lc}
\toprule
\multicolumn{2}{c}{Regional Style} \\
\midrule
Style & Count \\
\midrule
World & 86 \\
Japanese & 42 \\
Celtic & 36 \\
Middle Eastern & 33 \\
Indian & 22 \\
Latin American & 18 \\
Chinese & 17 \\
Carribean & 11 \\
South American & 7 \\
Appalachian & 6 \\
Slavic & 5 \\
African & 5 \\
Native American & 3 \\
Mediterranean & 2 \\
Cajun & 1 \\
\bottomrule
\end{tabular}
\end{minipage}
\hfill
\begin{minipage}{0.48\columnwidth}
\centering
\begin{tabular}{lc}
\toprule
\multicolumn{2}{c}{Mood} \\
\midrule
Mood & Count \\
\midrule
Energetic & 843 \\
Mellow & 617 \\
Dark & 541 \\
Aggressive & 530 \\
Chill & 478 \\
Happy & 376 \\
Funky & 334 \\
Suspenseful & 284 \\
Epic & 240 \\
Solemn & 181 \\
Jazzy & 179 \\
Quirky & 175 \\
Spooky & 117 \\
Sad & 88 \\
Other & 412 \\
\bottomrule
\end{tabular}
\end{minipage}

\end{table}

\begin{table}[t]
\centering
\caption{OverClocked ReMix Time Signatures}
\label{tab:time_signatures}
\begin{tabular}{lcccccc}
\toprule
Time Signature & 4/4 & 3/4 & 6/8 & 5/4 & 7/8 & 7/4 \\
\midrule
Count          & 844 & 63  & 53  & 11  & 5   & 3   \\
\bottomrule
\end{tabular}
\end{table}

For time signature identification, high scores reflect a skewed distribution toward 4/4 in the dataset illustrated in Table~\ref{tab:time_signatures}, which roughly matches the real-world distribution of time signatures. A model that always guesses 4/4 will therefore score well. Time signature is a particularly notable negative result: for every model, predictions are no more accurate when given the actual audio than when given random audio. Only Qwen3-Omni and Gemini~3 generate any label other than~4/4; Qwen3-Omni achieves the highest score because it correctly identifies most 4/4 cases and a handful of 3/4 cases. Gemini~3 generates a wider variety of time signatures (7/8, 9/8, etc.) but at the cost of mislabeling many true 4/4 cases. A model that always guesses 4/4 scores 0.869, matching or exceeding every model.
\vspace{-3mm}

\section{Conclusion}

We developed a new protocol for evaluation of LALMs, factual QA, and used this protocol to develop a suite of benchmarks, evaluating nine models across six tasks on three datasets. The protocol asks single-answer questions about audio, parses each model's free-form response into a closed vocabulary using a rule-bound extractor, and scores the result with precision, recall, and F1. Experimental analysis shows that the factual QA protocol clearly distinguishes between responses by LALMs conditioned on the correct audio versus a random audio baseline, in contrast to the free-form QA protocol studied in Section~\ref{sec:qa}. Alternative evaluation protocols such as LLM-as-a-judge and multiple-choice QA face their own challenges, which we discussed in Section~\ref{related_work}. We therefore see factual QA as a valuable addition to the toolbox for evaluation of LALMs. We do not claim factual QA exhausts what one might want to know about a LALM; it is, however, a concrete way to pose questions to these models whose answers can be checked, providing a new window into the capabilities of these models.

\section{AI Usage Statement}

We used an LLM for light copy-editing and proofreading of this document; otherwise all writing and analysis was performed by the authors. We used large language models (Gemini 3, GPT-4) extensively as part of our experiments, specifically within the evaluation pipeline that converts free-form model outputs into a canonical representation and related analysis steps as described in Section~\ref{sec:factuality}.

\section{Acknowledgments}
This work was supported in part by a research gift from SoundPatrol, Inc. We acknowledge generous support from a Cornell donor.

\bibliography{ISMIRtemplate}

@inproceedings{ltu-as,
  title={Joint audio and speech understanding},
  author={Gong, Yuan and Liu, Alexander H and Luo, Hongyin and Karlinsky, Leonid and Glass, James},
  booktitle={2023 IEEE Automatic Speech Recognition and Understanding Workshop (ASRU)},
  year={2023},
}

@inproceedings{bleu,
    title = "{B}leu: a Method for Automatic Evaluation of Machine Translation",
    author = "Papineni, Kishore  and
      Roukos, Salim  and
      Ward, Todd  and
      Zhu, Wei-Jing",
    editor = "Isabelle, Pierre  and
      Charniak, Eugene  and
      Lin, Dekang",
    booktitle = "Proceedings of the 40th Annual Meeting of the Association for Computational Linguistics",
    month = jul,
    year = "2002",
    address = "Philadelphia, Pennsylvania, USA",
    publisher = "Association for Computational Linguistics",
    url = "https://aclanthology.org/P02-1040/",
    doi = "10.3115/1073083.1073135",
    pages = "311--318"
}

@misc{bertscore,
      title={BERTScore: Evaluating Text Generation with BERT}, 
      author={Tianyi Zhang and Varsha Kishore and Felix Wu and Kilian Q. Weinberger and Yoav Artzi},
      year={2020},
      eprint={1904.09675},
      archivePrefix={arXiv},
      primaryClass={cs.CL},
      url={https://arxiv.org/abs/1904.09675}, 
}

@misc{cider,
      title={CIDEr: Consensus-based Image Description Evaluation}, 
      author={Ramakrishna Vedantam and C. Lawrence Zitnick and Devi Parikh},
      year={2015},
      eprint={1411.5726},
      archivePrefix={arXiv},
      primaryClass={cs.CV},
      url={https://arxiv.org/abs/1411.5726}, 
}

@misc{mu-llama,
      title={Music Understanding LLaMA: Advancing Text-to-Music Generation with Question Answering and Captioning}, 
      author={Shansong Liu and Atin Sakkeer Hussain and Chenshuo Sun and Ying Shan},
      year={2023},
      eprint={2308.11276},
      archivePrefix={arXiv},
      primaryClass={cs.SD},
      url={https://arxiv.org/abs/2308.11276}, 
}

@misc{imagebind-llm,
      title={ImageBind-LLM: Multi-modality Instruction Tuning}, 
      author={Jiaming Han and Renrui Zhang and Wenqi Shao and Peng Gao and Peng Xu and Han Xiao and Kaipeng Zhang and Chris Liu and Song Wen and Ziyu Guo and Xudong Lu and Shuai Ren and Yafei Wen and Xiaoxin Chen and Xiangyu Yue and Hongsheng Li and Yu Qiao},
      year={2023},
      eprint={2309.03905},
      archivePrefix={arXiv},
      primaryClass={cs.MM},
      url={https://arxiv.org/abs/2309.03905}, 
}

@misc{salmonn,
      title={SALMONN: Towards Generic Hearing Abilities for Large Language Models}, 
      author={Changli Tang and Wenyi Yu and Guangzhi Sun and Xianzhao Chen and Tian Tan and Wei Li and Lu Lu and Zejun Ma and Chao Zhang},
      year={2024},
      eprint={2310.13289},
      archivePrefix={arXiv},
      primaryClass={cs.SD},
      url={https://arxiv.org/abs/2310.13289}, 
}

@misc{llama,
      title={LLaMA: Open and Efficient Foundation Language Models}, 
      author={Hugo Touvron and Thibaut Lavril and Gautier Izacard and Xavier Martinet and Marie-Anne Lachaux and Timothée Lacroix and Baptiste Rozière and Naman Goyal and Eric Hambro and Faisal Azhar and Aurelien Rodriguez and Armand Joulin and Edouard Grave and Guillaume Lample},
      year={2023},
      eprint={2302.13971},
      archivePrefix={arXiv},
      primaryClass={cs.CL},
      url={https://arxiv.org/abs/2302.13971}, 
}

@misc{mert,
      title={MERT: Acoustic Music Understanding Model with Large-Scale Self-supervised Training}, 
      author={Yizhi Li and Ruibin Yuan and Ge Zhang and Yinghao Ma and Xingran Chen and Hanzhi Yin and Chenghao Xiao and Chenghua Lin and Anton Ragni and Emmanouil Benetos and Norbert Gyenge and Roger Dannenberg and Ruibo Liu and Wenhu Chen and Gus Xia and Yemin Shi and Wenhao Huang and Zili Wang and Yike Guo and Jie Fu},
      year={2024},
      eprint={2306.00107},
      archivePrefix={arXiv},
      primaryClass={cs.SD},
      url={https://arxiv.org/abs/2306.00107}, 
}

@misc{llark,
      title={LLark: A Multimodal Instruction-Following Language Model for Music}, 
      author={Josh Gardner and Simon Durand and Daniel Stoller and Rachel M. Bittner},
      year={2024},
      eprint={2310.07160},
      archivePrefix={arXiv},
      primaryClass={cs.SD},
      url={https://arxiv.org/abs/2310.07160}, 
}

@misc{llama-adapter,
      title={LLaMA-Adapter V2: Parameter-Efficient Visual Instruction Model}, 
      author={Peng Gao and Jiaming Han and Renrui Zhang and Ziyi Lin and Shijie Geng and Aojun Zhou and Wei Zhang and Pan Lu and Conghui He and Xiangyu Yue and Hongsheng Li and Yu Qiao},
      year={2023},
      eprint={2304.15010},
      archivePrefix={arXiv},
      primaryClass={cs.CV},
      url={https://arxiv.org/abs/2304.15010}, 
}

@misc{imagebind,
      title={ImageBind: One Embedding Space To Bind Them All}, 
      author={Rohit Girdhar and Alaaeldin El-Nouby and Zhuang Liu and Mannat Singh and Kalyan Vasudev Alwala and Armand Joulin and Ishan Misra},
      year={2023},
      eprint={2305.05665},
      archivePrefix={arXiv},
      primaryClass={cs.CV},
      url={https://arxiv.org/abs/2305.05665}, 
}

@inproceedings{tltr,
   title={Whisper-AT: Noise-Robust Automatic Speech Recognizers are Also Strong General Audio Event Taggers},
   booktitle={INTERSPEECH 2023},
   publisher={ISCA},
   author={Gong, Yuan and Khurana, Sameer and Karlinsky, Leonid and Glass, James},
   year={2023},
}

@misc{whisper,
      title={Robust Speech Recognition via Large-Scale Weak Supervision}, 
      author={Alec Radford and Jong Wook Kim and Tao Xu and Greg Brockman and Christine McLeavey and Ilya Sutskever},
      year={2022},
      eprint={2212.04356},
      archivePrefix={arXiv},
      primaryClass={eess.AS},
      url={https://arxiv.org/abs/2212.04356}, 
}

@misc{beats,
      title={BEATs: Audio Pre-Training with Acoustic Tokenizers}, 
      author={Sanyuan Chen and Yu Wu and Chengyi Wang and Shujie Liu and Daniel Tompkins and Zhuo Chen and Furu Wei},
      year={2022},
      eprint={2212.09058},
      archivePrefix={arXiv},
      primaryClass={eess.AS},
      url={https://arxiv.org/abs/2212.09058}, 
}

@misc{q-former,
      title={BLIP-2: Bootstrapping Language-Image Pre-training with Frozen Image Encoders and Large Language Models}, 
      author={Junnan Li and Dongxu Li and Silvio Savarese and Steven Hoi},
      year={2023},
      eprint={2301.12597},
      archivePrefix={arXiv},
      primaryClass={cs.CV},
      url={https://arxiv.org/abs/2301.12597}, 
}

@misc{musicnet,
      title={Learning Features of Music from Scratch}, 
      author={John Thickstun and Zaid Harchaoui and Sham Kakade},
      year={2017},
      eprint={1611.09827},
      archivePrefix={arXiv},
      primaryClass={stat.ML},
      url={https://arxiv.org/abs/1611.09827}, 
}

@misc{fma,
      title={FMA: A Dataset For Music Analysis}, 
      author={Michaël Defferrard and Kirell Benzi and Pierre Vandergheynst and Xavier Bresson},
      year={2017},
      eprint={1612.01840},
      archivePrefix={arXiv},
      primaryClass={cs.SD},
      url={https://arxiv.org/abs/1612.01840}, 
}

@misc{mumu-llama,
      title={MuMu-LLaMA: Multi-modal Music Understanding and Generation via Large Language Models}, 
      author={Shansong Liu and Atin Sakkeer Hussain and Qilong Wu and Chenshuo Sun and Ying Shan},
      year={2024},
      eprint={2412.06660},
      archivePrefix={arXiv},
      primaryClass={cs.SD},
      url={https://arxiv.org/abs/2412.06660}, 
}

@inproceedings{meteor,
  title={METEOR: An automatic metric for MT evaluation with improved correlation with human judgments},
  author={Banerjee, Satanjeev and Lavie, Alon},
  booktitle={Proceedings of the acl workshop on intrinsic and extrinsic evaluation measures for machine translation and/or summarization},
  pages={65--72},
  year={2005}
}

@inproceedings{rouge,
    title = "{ROUGE}: A Package for Automatic Evaluation of Summaries",
    author = "Lin, Chin-Yew",
    booktitle = "Text Summarization Branches Out",
    month = jul,
    year = "2004",
    address = "Barcelona, Spain",
    publisher = "Association for Computational Linguistics",
    url = "https://aclanthology.org/W04-1013/",
    pages = "74--81"
}

@inproceedings{muchomusic,
   title={MuChoMusic: Evaluating Music Understanding in Multimodal Audio-Language Models},
   author={Weck, Benno and Manco, Ilaria and Benetos, Emmanouil and Quinton, Elio and Fazekas, György and Bogdanov, Dmitry},
   booktitle = {Proceedings of the 25th International Society for Music Information Retrieval Conference (ISMIR)},
   year={2024}
}

@inproceedings{areyoureallylistening,
  title={Are you really listening? Boosting Perceptual Awareness in Music-QA Benchmarks},
  author={Yongyi Zang and Sean O'Brien and Taylor Berg-Kirkpatrick and Julian McAuley and Zachary Novack},
  booktitle = {Proceedings of the 26th International Society for Music Information Retrieval Conference (ISMIR)},
  year={2025}
}

@article{captioning,
  title={Do Captioning Metrics Reflect Music Semantic Alignment?},
  author={Lee, Jinwoo and Lee, Kyogu},
  journal={arXiv preprint arXiv:2411.11692},
  year={2024}
}

@misc{lm-judge,
      title={Judging LLM-as-a-Judge with MT-Bench and Chatbot Arena}, 
      author={Lianmin Zheng and Wei-Lin Chiang and Ying Sheng and Siyuan Zhuang and Zhanghao Wu and Yonghao Zhuang and Zi Lin and Zhuohan Li and Dacheng Li and Eric P. Xing and Hao Zhang and Joseph E. Gonzalez and Ion Stoica},
      year={2023},
      eprint={2306.05685},
      archivePrefix={arXiv},
      primaryClass={cs.CL},
      url={https://arxiv.org/abs/2306.05685}, 
}

@misc{qwen2-audio,
      title={Qwen2-Audio Technical Report}, 
      author={Yunfei Chu and Jin Xu and Qian Yang and Haojie Wei and Xipin Wei and Zhifang Guo and Yichong Leng and Yuanjun Lv and Jinzheng He and Junyang Lin and Chang Zhou and Jingren Zhou},
      year={2024},
      eprint={2407.10759},
      archivePrefix={arXiv},
      primaryClass={eess.AS},
      url={https://arxiv.org/abs/2407.10759}, 
}

@misc{qwen3-omni,
      title={Qwen3-Omni Technical Report}, 
      author={Jin Xu and Zhifang Guo and Hangrui Hu and Yunfei Chu and Xiong Wang and Jinzheng He and Yuxuan Wang and Xian Shi and Ting He and Xinfa Zhu and Yuanjun Lv and Yongqi Wang and Dake Guo and He Wang and Linhan Ma and Pei Zhang and Xinyu Zhang and Hongkun Hao and Zishan Guo and Baosong Yang and Bin Zhang and Ziyang Ma and Xipin Wei and Shuai Bai and Keqin Chen and Xuejing Liu and Peng Wang and Mingkun Yang and Dayiheng Liu and Xingzhang Ren and Bo Zheng and Rui Men and Fan Zhou and Bowen Yu and Jianxin Yang and Le Yu and Jingren Zhou and Junyang Lin},
      year={2025},
      eprint={2509.17765},
      archivePrefix={arXiv},
      primaryClass={cs.CL},
      url={https://arxiv.org/abs/2509.17765}, 
}

@misc{audio-flamingo3,
      title={Audio Flamingo 3: Advancing Audio Intelligence with Fully Open Large Audio Language Models}, 
      author={Arushi Goel and Sreyan Ghosh and Jaehyeon Kim and Sonal Kumar and Zhifeng Kong and Sang-gil Lee and Chao-Han Huck Yang and Ramani Duraiswami and Dinesh Manocha and Rafael Valle and Bryan Catanzaro},
      year={2025},
      eprint={2507.08128},
      archivePrefix={arXiv},
      primaryClass={cs.SD},
      url={https://arxiv.org/abs/2507.08128}, 
}

@misc{music-flamingo,
      title={Music Flamingo: Scaling Music Understanding in Audio Language Models}, 
      author={Sreyan Ghosh and Arushi Goel and Lasha Koroshinadze and Sang-gil Lee and Zhifeng Kong and Joao Felipe Santos and Ramani Duraiswami and Dinesh Manocha and Wei Ping and Mohammad Shoeybi and Bryan Catanzaro},
      year={2025},
      eprint={2511.10289},
      archivePrefix={arXiv},
      primaryClass={eess.AS},
      url={https://arxiv.org/abs/2511.10289}, 
}

@misc{audio-flamingo2,
      title={Audio Flamingo 2: An Audio-Language Model with Long-Audio Understanding and Expert Reasoning Abilities}, 
      author={Sreyan Ghosh and Zhifeng Kong and Sonal Kumar and S Sakshi and Jaehyeon Kim and Wei Ping and Rafael Valle and Dinesh Manocha and Bryan Catanzaro},
      year={2025},
      eprint={2503.03983},
      archivePrefix={arXiv},
      primaryClass={cs.SD},
      url={https://arxiv.org/abs/2503.03983}, 
}

@misc{audio-flamingo,
      title={Audio Flamingo: A Novel Audio Language Model with Few-Shot Learning and Dialogue Abilities}, 
      author={Zhifeng Kong and Arushi Goel and Rohan Badlani and Wei Ping and Rafael Valle and Bryan Catanzaro},
      year={2024},
      eprint={2402.01831},
      archivePrefix={arXiv},
      primaryClass={cs.SD},
      url={https://arxiv.org/abs/2402.01831}, 
}

@misc{clotho,
      title={Clotho: An Audio Captioning Dataset}, 
      author={Konstantinos Drossos and Samuel Lipping and Tuomas Virtanen},
      year={2019},
      eprint={1910.09387},
      archivePrefix={arXiv},
      primaryClass={cs.SD},
      url={https://arxiv.org/abs/1910.09387}, 
}

@inproceedings{audiocaps,
    title = "{A}udio{C}aps: Generating Captions for Audios in The Wild",
    author = "Kim, Chris Dongjoo  and
      Kim, Byeongchang  and
      Lee, Hyunmin  and
      Kim, Gunhee",
    editor = "Burstein, Jill  and
      Doran, Christy  and
      Solorio, Thamar",
    booktitle = "Proceedings of the 2019 Conference of the North {A}merican Chapter of the Association for Computational Linguistics: Human Language Technologies, Volume 1 (Long and Short Papers)",
    month = jun,
    year = "2019",
    address = "Minneapolis, Minnesota",
    publisher = "Association for Computational Linguistics",
    url = "https://aclanthology.org/N19-1011/",
    doi = "10.18653/v1/N19-1011",
    pages = "119--132"
}

@misc{mmau,
      title={MMAU: A Massive Multi-Task Audio Understanding and Reasoning Benchmark}, 
      author={S Sakshi and Utkarsh Tyagi and Sonal Kumar and Ashish Seth and Ramaneswaran Selvakumar and Oriol Nieto and Ramani Duraiswami and Sreyan Ghosh and Dinesh Manocha},
      year={2024},
      eprint={2410.19168},
      archivePrefix={arXiv},
      primaryClass={eess.AS},
      url={https://arxiv.org/abs/2410.19168}, 
}

@misc{robustness,
      title={Robustness assessment of large audio language models in multiple-choice evaluation}, 
      author={Fernando López and Santosh Kesiraju and Jordi Luque},
      year={2025},
      eprint={2510.04584},
      archivePrefix={arXiv},
      primaryClass={cs.CL},
      url={https://arxiv.org/abs/2510.04584}, 
}

@misc{qwen-audio,
      title={Qwen-Audio: Advancing Universal Audio Understanding via Unified Large-Scale Audio-Language Models}, 
      author={Yunfei Chu and Jin Xu and Xiaohuan Zhou and Qian Yang and Shiliang Zhang and Zhijie Yan and Chang Zhou and Jingren Zhou},
      year={2023},
      eprint={2311.07919},
      archivePrefix={arXiv},
      primaryClass={eess.AS},
      url={https://arxiv.org/abs/2311.07919}, 
}

@misc{qwen2.5-omni,
      title={Qwen2.5-Omni Technical Report}, 
      author={Jin Xu and Zhifang Guo and Jinzheng He and Hangrui Hu and Ting He and Shuai Bai and Keqin Chen and Jialin Wang and Yang Fan and Kai Dang and Bin Zhang and Xiong Wang and Yunfei Chu and Junyang Lin},
      year={2025},
      eprint={2503.20215},
      archivePrefix={arXiv},
      primaryClass={cs.CL},
      url={https://arxiv.org/abs/2503.20215}, 
}

@misc{gemini3,
  author       = {{Google DeepMind}},
  title        = {Gemini 3 Pro Model Card},
  year         = {2025},
  howpublished = {\url{https://deepmind.google/models/model-cards/gemini-3-pro/}},
  note         = {Accessed: 2026-04-26}
}

@misc{gemini3vertex,
  author       = {{Google Cloud}},
  title        = {Gemini 3 Pro --- Vertex AI Documentation},
  year         = {2026},
  howpublished = {\url{https://cloud.google.com/vertex-ai/generative-ai/docs/models/gemini/3-pro}},
  note         = {Accessed: 2026-04-26}
}

@misc{fever,
      title={FEVER: a large-scale dataset for Fact Extraction and VERification}, 
      author={James Thorne and Andreas Vlachos and Christos Christodoulopoulos and Arpit Mittal},
      year={2018},
      eprint={1803.05355},
      archivePrefix={arXiv},
      primaryClass={cs.CL},
      url={https://arxiv.org/abs/1803.05355}, 
}

@misc{factscore,
      title={FActScore: Fine-grained Atomic Evaluation of Factual Precision in Long Form Text Generation}, 
      author={Sewon Min and Kalpesh Krishna and Xinxi Lyu and Mike Lewis and Wen-tau Yih and Pang Wei Koh and Mohit Iyyer and Luke Zettlemoyer and Hannaneh Hajishirzi},
      year={2023},
      eprint={2305.14251},
      archivePrefix={arXiv},
      primaryClass={cs.CL},
      url={https://arxiv.org/abs/2305.14251}, 
}

@misc{qwen,
      title={Qwen Technical Report}, 
      author={Jinze Bai and Shuai Bai and Yunfei Chu and Zeyu Cui and Kai Dang and Xiaodong Deng and Yang Fan and Wenbin Ge and Yu Han and Fei Huang and Binyuan Hui and Luo Ji and Mei Li and Junyang Lin and Runji Lin and Dayiheng Liu and Gao Liu and Chengqiang Lu and Keming Lu and Jianxin Ma and Rui Men and Xingzhang Ren and Xuancheng Ren and Chuanqi Tan and Sinan Tan and Jianhong Tu and Peng Wang and Shijie Wang and Wei Wang and Shengguang Wu and Benfeng Xu and Jin Xu and An Yang and Hao Yang and Jian Yang and Shusheng Yang and Yang Yao and Bowen Yu and Hongyi Yuan and Zheng Yuan and Jianwei Zhang and Xingxuan Zhang and Yichang Zhang and Zhenru Zhang and Chang Zhou and Jingren Zhou and Xiaohuan Zhou and Tianhang Zhu},
      year={2023},
      eprint={2309.16609},
      archivePrefix={arXiv},
      primaryClass={cs.CL},
      url={https://arxiv.org/abs/2309.16609}, 
}

\appendix

\section{Details on Evaluated LALMs}
\label{model_details}

We evaluate eight open-source Large Audio Language Models (LALMs) and one
closed-source baseline. The first four (LTU-AS, MU-LLaMA, LLaMA-Adapter,
SALMONN) constitute the set of LALMs evaluated in prior work on music
question answering; the remaining four (Qwen2-Audio, Qwen3-Omni,
Audio-Flamingo~3, Music-Flamingo) are more recent systems with substantially
stronger language-model backbones.

\subsection{Original Four LALMs}

The LTU-AS model~\cite{ltu-as} is a general-purpose audio language model
finetuned from LLaMA-7B~\cite{llama} for a variety of QA tasks including
speech, music, and sound effects. LTU-AS processes audio inputs tokenized
using Whisper~\cite{whisper} and TLTR~\cite{tltr} encoders, and is trained on
the Open-ASQA dataset developed by the LTU-AS authors. This dataset
aggregates samples from multiple sources, including AS-Strong, AudioSet,
VGGSound, FSD50K, AudioCaps, FreeSound, Clotho, SoundBible, IEMOCAP,
LibriTTS, VoxCeleb2, MOSEI, and FMA.

LLaMA-Adapter~\cite{llama-adapter} is a general-purpose multimodal language
model finetuned from LLaMA-7B and designed to target a variety of
modalities including audio and, specifically, music. In this paper, we use
ImageBind-LLM~\cite{imagebind-llm}, one of the latest variants of
LLaMA-Adapter. ImageBind-LLM tokenizes text, audio, video, and images into a
shared ImageBind embedding space~\cite{imagebind}. ImageBind itself is
trained on a combination of datasets, including AudioSet, ESC (5-fold),
Clotho, AudioCaps, and VGGSound.

MU-LLaMA~\cite{mu-llama} is designed specifically for music captioning and
music question--answering tasks. It follows the LLaMA-Adapter
architecture derived from LLaMA-7B, but replaces Whisper-based audio
tokenization with MERT embeddings~\cite{mert}. MU-LLaMA is trained,
fine-tuned, and evaluated using the MusicQA dataset developed by its authors.

SALMONN~\cite{salmonn} is a general-purpose audio language model that
processes audio tokenized using Whisper and BEATs~\cite{beats}
representations fused according to the Q-Former architecture~\cite{q-former}.
While multiple variants of SALMONN exist, including newer versions
specialized for video and speech, in our experiments we select the original
version of SALMONN to maintain focus on audio modeling. SALMONN is derived
from the Vicuna-7B fine-tuned variant of LLaMA-2 7B~\cite{lm-judge}. It is
trained on a diverse set of audio datasets, including LibriSpeech,
GigaSpeech, CoVoST2-En2Zh, AudioCaps, Clotho, IEMOCAP, MusicCaps, LibriMix,
VoxCeleb1, WavCaps, MillionSong, and MusicNet.

\subsection{Recent LALMs}

Qwen2-Audio~\cite{qwen2-audio} is a general-purpose audio language model
that processes audio tokenized using a Whisper-large-v3 encoder, which is
fed into the Qwen-7B language model backbone~\cite{qwen}. The full model
contains roughly 8.2B parameters. Audio inputs are pre-processed into
128-channel mel-spectrograms and pooled to roughly one frame per 40\,ms.
Qwen2-Audio is trained in three stages---pretraining on natural-language
prompts across diverse audio tasks, supervised instruction tuning, and
direct preference optimization---and supports two interaction modes
(``audio analysis'' and ``voice chat'') that share the same weights. The
training corpus spans speech, sound, and music. We use the publicly
released \texttt{Qwen2-Audio-7B-Instruct} checkpoint.

Qwen3-Omni~\cite{qwen3-omni} is the omni-modal successor to Qwen2-Audio,
unifying text, image, audio, and video inputs in a single end-to-end
model. It adopts a Mixture-of-Experts (MoE) Thinker--Talker architecture,
where the Thinker handles multimodal reasoning and text generation while
the Talker handles streaming speech synthesis; we evaluate only the
Thinker component, which produces text outputs from audio inputs.
Qwen3-Omni replaces the Whisper-based audio encoder used in Qwen2-Audio
with a new \emph{Audio Transformer} (AuT) encoder trained from scratch on
roughly 20 million hours of supervised audio, producing general-purpose
audio representations at a token rate of 12.5\,Hz. We use the
\texttt{Qwen3-Omni-30B-A3B-Instruct} checkpoint.

Audio-Flamingo~3~\cite{audio-flamingo3} is a fully open large
audio-language model that aims to handle speech, sound, and music within
a single architecture. It consists of a unified audio encoder
(AF-Whisper, derived from Whisper-large-v3 and finetuned for joint
speech/sound/music representation), an MLP-based audio adaptor, and a
Qwen2.5-7B language model backbone in a LLaVA-style configuration. The
audio encoder operates on 30-second windows and the model can ingest up
to 10 minutes of audio per sample (windows beyond this are truncated).
Audio-Flamingo~3 is trained on roughly 50M audio--text pairs through a
five-stage curriculum that includes alignment pretraining, encoder
tuning, full finetuning on the AudioSkills-XL and LongAudio-XL datasets,
chain-of-thought training, and chat finetuning. We use the
\texttt{audio-flamingo-3-hf} checkpoint released by NVIDIA.

Music-Flamingo~\cite{music-flamingo} is a music-specialized model that
inherits the Audio-Flamingo~3 architecture (AF-Whisper encoder, LLaVA
fusion, Qwen2.5-7B backbone, $\sim$8B parameters total) and finetunes it
on music-specific data. It introduces Rotary Time Embeddings (RoTE) to
attach absolute timestamps to audio tokens, supports audio inputs up to
20 minutes, and is trained on MF-Skills, a music dataset spanning $\sim$2M
full-length songs across more than 100 genres and cultural contexts with
captions covering harmony, structure, timbre, and lyrics. Training
includes a chain-of-thought stage on MF-Think followed by GRPO-based
reinforcement learning with music-theory-aware rewards. Music-Flamingo
is the only model in our evaluation explicitly trained at scale on
full-length music audio, and we treat it as the strongest open-source
music-specialized LALM at the time of writing. We use the
\texttt{music-flamingo-hf} checkpoint released by NVIDIA.

\subsection{Closed-Source Baseline}

Gemini~3~\cite{gemini3} is Google's third-generation multimodal
foundation model, released in November~2025. It is a closed-source
proprietary model accessible only through Google's APIs; architectural
details, parameter counts, audio encoder specifics, and training data
have not been publicly disclosed. Gemini~3 natively accepts text,
images, audio, video, and code as input within a context window of up to
1M tokens. We include Gemini~3 in our evaluation as a strong
non-music-specialized reference: although Google notes that ``Gemini~3
Pro models aren't designed around prioritizing audio understanding''~\cite{gemini3vertex},
the model's overall scale and multimodal training make it a useful upper
reference point for what a frontier general-purpose model can recover
from music audio without targeted finetuning. We query the
\texttt{Gemini-3-pro} model via the Google AI Studio API; all other
inference settings (temperature, thinking level, etc.) are kept at their
defaults.

\subsection{Models Excluded From This Study}
 
We do not separately benchmark the earlier versions of model families
already represented above by their most recent release. This includes
Qwen-Audio~\cite{qwen-audio} (the predecessor to Qwen2-Audio) and
Qwen2.5-Omni~\cite{qwen2.5-omni} (the predecessor to Qwen3-Omni), as
well as Audio Flamingo~\cite{audio-flamingo} and Audio
Flamingo~2~\cite{audio-flamingo2} (the predecessors to Audio
Flamingo~3). In each case, the newer release uses a stronger LLM
backbone, a stronger audio encoder, and substantially more training
data than its predecessor, and outperforms it across the relevant
benchmarks reported by the authors; we therefore treat the newer
release as representative of the family for our purposes.
 
Beyond these family predecessors, two other music-focused models warrant
mention. LLark~\cite{llark} enables more detailed music captioning on
longer paragraphs, but has not released checkpoints for replication.
MuMu-LLaMA~\cite{mumu-llama} is another recent music-focused model,
though the publicly released checkpoint is corrupted and cannot be
systematically tested. As more usable checkpoints are released, the
evaluation framework established in this paper provides a foundation for
assessing factual correctness in music--language modeling.
 
\section{Parsing Protocol}
\label{keyword_extraction}

\subsection{Parsing Rules}
\label{keyword_extraction_rules}

We define the following rules for factual output parsing, which are
consistently applied across all tasks and categories:

\begin{enumerate}
    \item \textbf{Exact mention requirement:} Only terms that explicitly
    appear in the model's output are returned.

    \item \textbf{No guessing:} Implicit inference or contextual
    associations (e.g., ``jazz'' inferred from ``swing rhythm'') are
    excluded.

    \item \textbf{Comparatives:} For comparative statements (``X more than
    Y''), only the preferred entity (X) is retained.

    \item \textbf{Deduplication and ordering:} Duplicate mentions are
    removed, while the order of first appearance is preserved.

    \item \textbf{Output format:} Results are returned as a comma-separated
    list, e.g., \emph{rock, jazz, classical}.

    \item \textbf{Empty case:} If no relevant term is mentioned, the output
    is an empty string.

    \item \textbf{Canonical form:} All terms are normalized to simplified
    canonical forms (e.g., ``J.S. Bach'' $\rightarrow$ ``Bach'',
    ``Acoustic Grand Piano'' $\rightarrow$ ``piano'').

    \item \textbf{No stylistic inclusion:} Descriptions of mood or affect
    without explicit mention (e.g., ``a jazz-like feeling'') are ignored.
\end{enumerate}

For tasks whose canonical vocabulary admits genuine semantic overlap among labels --- specifically mood and regional style on OverClocked ReMix--- we adopt a complementary \emph{label-matching} variant of the protocol. Rather than asking the parser to extract whatever terms the model uttered and then compare them to the ground-truth label, we present the parser with the full closed list of dataset labels and ask it to select the labels best reflected by the model's response. For example, a model output of \emph{Egyptian} is matched to \emph{Middle Eastern}, and \emph{upbeat} is matched to whichever of {\emph{energetic}, \emph{happy}, \ldots} the parser judges most appropriate. The eight rules above otherwise apply unchanged: extraction is still grounded in what the model explicitly said, and inference beyond the response is still disallowed. Label matching is not used for instrumentation, composer, genre, or time signature, where the canonical forms are unambiguous and exact-mention extraction suffices.

\subsection{Parser Robustness: GPT-4.1 vs.\ Gemini 3}
\label{parser_robustness}

A reasonable concern with our protocol is that the choice of parser LLM
might bias the reported factuality scores: if GPT-4.1-mini systematically
over- or under-extracts certain labels, the resulting numbers could reflect
parser idiosyncrasies rather than LALM behavior. To test the sensitivity of
our framework to this choice, we re-ran the output parsing step using
Gemini in place of GPT-4.1-mini, on two of the OverClocked ReMix tasks (mood and
regional style), and recomputed all factuality metrics. Tables
~\ref{tab:parser-mood} and ~\ref{tab:regional} report the side-by-side
comparison.

Across both tasks and all five models, the absolute differences between
GPT-parsed and Gemini-parsed F1 scores are small (typically under 0.02 in
absolute value), and the relative ranking of LALMs is preserved in every
case. This suggests that the label-parsing step is largely robust to
the specific parser used, provided the prompt enforces the strict rules
listed in Appendix~\ref{keyword_extraction_rules}. We use GPT-4.1-mini as
our default parser in the main results for reproducibility, but practitioners
can substitute another competent instruction-following LLM with minimal
impact on conclusions.

\begin{table*}[t]
\centering
\caption{Parser comparison for mood classification on OverClocked ReMix: GPT-parsed
vs.\ Gemini-parsed labels. Differences are small relative to the
gap between Correct and Wrong columns, indicating that factuality results
are not sensitive to the specific parser used.}
\label{tab:parser-mood}
\resizebox{\textwidth}{!}{%
\begin{tabular}{l cc cc cc cc cc cc cc cc cc cc}
\toprule
& \multicolumn{4}{c}{Qwen2-Audio} & \multicolumn{4}{c}{Qwen3-Omni} & \multicolumn{4}{c}{Audio-Flamingo 3} & \multicolumn{4}{c}{Music-Flamingo} & \multicolumn{4}{c}{Gemini 3} \\
\cmidrule(lr){2-5} \cmidrule(lr){6-9} \cmidrule(lr){10-13} \cmidrule(lr){14-17} \cmidrule(lr){18-21}
& \multicolumn{2}{c}{GPT-Parsed} & \multicolumn{2}{c}{Gemini-Parsed} & \multicolumn{2}{c}{GPT-Parsed} & \multicolumn{2}{c}{Gemini-Parsed} & \multicolumn{2}{c}{GPT-Parsed} & \multicolumn{2}{c}{Gemini-Parsed} & \multicolumn{2}{c}{GPT-Parsed} & \multicolumn{2}{c}{Gemini-Parsed} & \multicolumn{2}{c}{GPT-Parsed} & \multicolumn{2}{c}{Gemini-Parsed} \\
\cmidrule(lr){2-3} \cmidrule(lr){4-5} \cmidrule(lr){6-7} \cmidrule(lr){8-9} \cmidrule(lr){10-11} \cmidrule(lr){12-13} \cmidrule(lr){14-15} \cmidrule(lr){16-17} \cmidrule(lr){18-19} \cmidrule(lr){20-21}
& Correct & Wrong & Correct & Wrong & Correct & Wrong & Correct & Wrong & Correct & Wrong & Correct & Wrong & Correct & Wrong & Correct & Wrong & Correct & Wrong & Correct & Wrong \\
\midrule
Precision & 0.304 & 0.155 & 0.287 & 0.148 & 0.247 & 0.115 & 0.221 & 0.115 & 0.356 & 0.158 & 0.341 & 0.149 & 0.284 & 0.131 & 0.249 & 0.120 & 0.260 & 0.146 & 0.244 & 0.140 \\
Recall    & 0.263 & 0.133 & 0.261 & 0.133 & 0.487 & 0.226 & 0.530 & 0.275 & 0.352 & 0.156 & 0.353 & 0.154 & 0.324 & 0.150 & 0.347 & 0.167 & 0.403 & 0.226 & 0.440 & 0.253 \\
F1-Score  & 0.282 & 0.143 & 0.273 & 0.141 & 0.328 & 0.152 & 0.312 & 0.162 & 0.354 & 0.157 & 0.347 & 0.152 & 0.303 & 0.140 & 0.290 & 0.140 & 0.316 & 0.177 & 0.314 & 0.180 \\
\bottomrule
\end{tabular}%
}
\end{table*}

\section{Full Factuality Results}
\label{full_factuality_results}

\subsection{Results}

This appendix reports the complete factuality results across all
benchmarks, models, and the two main prompting conditions used in the
paper: an \emph{open} prompt that asks the LALM to answer freely, and a
\emph{list} prompt that explicitly enumerates the candidate labels in the
question. The main paper highlights selected entries from these tables;
we provide the full numbers here. Tables are organized by task:
instrumentation on MusicNet
(Tables~\ref{tab:instr-open}--\ref{tab:instr-list}), composer
classification on MusicNet
(Tables~\ref{tab:comp-open}--\ref{tab:comp-list}), genre classification
on FMA (Tables~\ref{tab:genre-open}--\ref{tab:genre-list}), and the
three OverClocked ReMix tasks of regional style (Table~\ref{tab:regional}), mood
(Table~\ref{tab:mood}), and time signature (Table~\ref{tab:timesig}).

\begin{table*}[t]
\centering
\caption{Instrument recognition on MusicNet. Prompt: ``Which instruments
are used in this piece of music?''}
\label{tab:instr-open}
\resizebox{\textwidth}{!}{%
\begin{tabular}{l cc cc cc cc cc cc cc cc}
\toprule
& \multicolumn{2}{c}{LTU-AS} & \multicolumn{2}{c}{MU-LLaMA} & \multicolumn{2}{c}{LLaMA-Adapter} & \multicolumn{2}{c}{Qwen2-Audio} & \multicolumn{2}{c}{Qwen3-Omni} & \multicolumn{2}{c}{Audio-Flamingo 3} & \multicolumn{2}{c}{Music-Flamingo} & \multicolumn{2}{c}{Gemini 3} \\
\cmidrule(lr){2-3} \cmidrule(lr){4-5} \cmidrule(lr){6-7} \cmidrule(lr){8-9} \cmidrule(lr){10-11} \cmidrule(lr){12-13} \cmidrule(lr){14-15} \cmidrule(lr){16-17}
& Correct & Wrong & Correct & Wrong & Correct & Wrong & Correct & Wrong & Correct & Wrong & Correct & Wrong & Correct & Wrong & Correct & Wrong \\
\midrule
Precision & 0.534 & 0.364 & 0.367 & 0.266 & 0.488 & 0.350 & 0.893 & 0.420 & 0.877 & 0.408 & 0.692 & 0.376 & 0.849 & 0.454 & 0.778 & 0.457 \\
Recall    & 0.461 & 0.314 & 0.582 & 0.422 & 0.676 & 0.486 & 0.520 & 0.244 & 0.633 & 0.294 & 0.767 & 0.417 & 0.698 & 0.373 & 0.529 & 0.311 \\
F1-Score  & 0.495 & 0.337 & 0.450 & 0.326 & 0.567 & 0.407 & 0.657 & 0.309 & 0.735 & 0.342 & 0.728 & 0.396 & 0.766 & 0.409 & 0.630 & 0.370 \\
\bottomrule
\end{tabular}%
}
\end{table*}

\begin{table*}[t]
\centering
\caption{Instrument recognition on MusicNet. Prompt: ``Among Acoustic Grand
Piano, Bassoon, Cello, Clarinet, Contrabass, Drums, Flute, French Horn,
Harpsichord, Oboe, Pizzicato Strings, Viola, and Violin, which instruments
are used in this piece of music?''}
\label{tab:instr-list}
\resizebox{\textwidth}{!}{%
\begin{tabular}{l cc cc cc cc cc cc cc cc}
\toprule
& \multicolumn{2}{c}{LTU-AS} & \multicolumn{2}{c}{MU-LLaMA} & \multicolumn{2}{c}{LLaMA-Adapter} & \multicolumn{2}{c}{Qwen2-Audio} & \multicolumn{2}{c}{Qwen3-Omni} & \multicolumn{2}{c}{Audio-Flamingo 3} & \multicolumn{2}{c}{Music-Flamingo} & \multicolumn{2}{c}{Gemini 3} \\
\cmidrule(lr){2-3} \cmidrule(lr){4-5} \cmidrule(lr){6-7} \cmidrule(lr){8-9} \cmidrule(lr){10-11} \cmidrule(lr){12-13} \cmidrule(lr){14-15} \cmidrule(lr){16-17}
& Correct & Wrong & Correct & Wrong & Correct & Wrong & Correct & Wrong & Correct & Wrong & Correct & Wrong & Correct & Wrong & Correct & Wrong \\
\midrule
Precision & 0.155 & 0.149 & 0.199 & 0.172 & 0.164 & 0.164 & 0.175 & 0.167 & 0.766 & 0.406 & 0.236 & 0.193 & 0.237 & 0.187 & 0.650 & 0.330 \\
Recall    & 0.714 & 0.689 & 0.773 & 0.668 & 0.923 & 0.920 & 0.812 & 0.776 & 0.760 & 0.403 & 0.881 & 0.721 & 0.864 & 0.682 & 0.582 & 0.295 \\
F1-Score  & 0.254 & 0.245 & 0.316 & 0.273 & 0.279 & 0.278 & 0.289 & 0.275 & 0.763 & 0.403 & 0.372 & 0.305 & 0.371 & 0.293 & 0.614 & 0.312 \\
\bottomrule
\end{tabular}%
}
\end{table*}

\begin{table*}[t]
\centering
\caption{Composer classification on MusicNet. Prompt: ``Which classical
composer's work does this sound like?''}
\label{tab:comp-open}
\resizebox{\textwidth}{!}{%
\begin{tabular}{l cc cc cc cc cc cc cc}
\toprule
& \multicolumn{2}{c}{LTU-AS} & \multicolumn{2}{c}{MU-LLaMA} & \multicolumn{2}{c}{LLaMA-Adapter} & \multicolumn{2}{c}{Qwen2-Audio} & \multicolumn{2}{c}{Qwen3-Omni} & \multicolumn{2}{c}{Music-Flamingo} & \multicolumn{2}{c}{Gemini 3} \\
\cmidrule(lr){2-3} \cmidrule(lr){4-5} \cmidrule(lr){6-7} \cmidrule(lr){8-9} \cmidrule(lr){10-11} \cmidrule(lr){12-13} \cmidrule(lr){14-15}
& Correct & Wrong & Correct & Wrong & Correct & Wrong & Correct & Wrong & Correct & Wrong & Correct & Wrong & Correct & Wrong \\
\midrule
Precision & 0.333 & 0.337 & 0.226 & 0.235 & 0.281 & 0.241 & 0.160 & 0.040 & 0.160 & 0.046 & 0.469 & 0.293 & 0.042 & 0.018 \\
Recall    & 0.285 & 0.288 & 0.618 & 0.642 & 0.506 & 0.433 & 0.177 & 0.044 & 0.177 & 0.051 & 0.471 & 0.294 & 0.066 & 0.028 \\
F1-Score  & 0.307 & 0.310 & 0.331 & 0.344 & 0.361 & 0.310 & 0.168 & 0.042 & 0.168 & 0.048 & 0.470 & 0.294 & 0.052 & 0.022 \\
\bottomrule
\end{tabular}%
}
\end{table*}

\begin{table*}[t]
\centering
\caption{Composer classification on MusicNet. Prompt: ``Among Bach,
Beethoven, Brahms, Cambini, Dvorak, Faure, Haydn, Mozart, Ravel, and
Schubert, whose style does this piece of music sound like?''}
\label{tab:comp-list}
\resizebox{\textwidth}{!}{%
\begin{tabular}{l cc cc cc cc cc cc cc}
\toprule
& \multicolumn{2}{c}{LTU-AS} & \multicolumn{2}{c}{MU-LLaMA} & \multicolumn{2}{c}{LLaMA-Adapter} & \multicolumn{2}{c}{Qwen2-Audio} & \multicolumn{2}{c}{Qwen3-Omni} & \multicolumn{2}{c}{Music-Flamingo} & \multicolumn{2}{c}{Gemini 3} \\
\cmidrule(lr){2-3} \cmidrule(lr){4-5} \cmidrule(lr){6-7} \cmidrule(lr){8-9} \cmidrule(lr){10-11} \cmidrule(lr){12-13} \cmidrule(lr){14-15}
& Correct & Wrong & Correct & Wrong & Correct & Wrong & Correct & Wrong & Correct & Wrong & Correct & Wrong & Correct & Wrong \\
\midrule
Precision & 0.152 & 0.133 & 0.153 & 0.140 & 0.152 & 0.137 & 0.152 & 0.096 & 0.152 & 0.085 & 0.204 & 0.201 & 0.056 & 0.046 \\
Recall    & 0.558 & 0.488 & 0.294 & 0.270 & 0.561 & 0.506 & 0.408 & 0.257 & 0.408 & 0.229 & 0.204 & 0.201 & 0.079 & 0.064 \\
F1-Score  & 0.238 & 0.209 & 0.201 & 0.185 & 0.239 & 0.216 & 0.221 & 0.140 & 0.221 & 0.124 & 0.204 & 0.201 & 0.066 & 0.053 \\
\bottomrule
\end{tabular}%
}
\end{table*}

\begin{table*}[t]
\centering
\caption{Genre classification on FMA. Prompt: ``What genre does this piece
of music fall under?''}
\label{tab:genre-open}
\resizebox{\textwidth}{!}{%
\begin{tabular}{l cc cc cc cc cc cc}
\toprule
& \multicolumn{2}{c}{MU-LLaMA} & \multicolumn{2}{c}{LLaMA-Adapter} & \multicolumn{2}{c}{SALMONN} & \multicolumn{2}{c}{Qwen2-Audio} & \multicolumn{2}{c}{Qwen3-Omni} & \multicolumn{2}{c}{Gemini 3} \\
\cmidrule(lr){2-3} \cmidrule(lr){4-5} \cmidrule(lr){6-7} \cmidrule(lr){8-9} \cmidrule(lr){10-11} \cmidrule(lr){12-13}
& Correct & Wrong & Correct & Wrong & Correct & Wrong & Correct & Wrong & Correct & Wrong & Correct & Wrong \\
\midrule
Precision & 0.256 & 0.102 & 0.334 & 0.081 & 0.293 & 0.084 & 0.445 & 0.097 & 0.216 & 0.065 & 0.219 & 0.071 \\
Recall    & 0.291 & 0.115 & 0.342 & 0.083 & 0.388 & 0.111 & 0.364 & 0.079 & 0.433 & 0.130 & 0.367 & 0.120 \\
F1-Score  & 0.272 & 0.108 & 0.338 & 0.082 & 0.334 & 0.096 & 0.401 & 0.087 & 0.289 & 0.086 & 0.274 & 0.089 \\
\bottomrule
\end{tabular}%
}
\end{table*}

\begin{table*}[t]
\centering
\caption{Genre classification on FMA. Prompt: ``Among hip-hop, pop, folk,
experimental, rock, international, electronic, instrumental, what is the
genre of this song?''}
\label{tab:genre-list}
\resizebox{\textwidth}{!}{%
\begin{tabular}{l cc cc cc cc cc cc}
\toprule
& \multicolumn{2}{c}{MU-LLaMA} & \multicolumn{2}{c}{LLaMA-Adapter} & \multicolumn{2}{c}{SALMONN} & \multicolumn{2}{c}{Qwen2-Audio} & \multicolumn{2}{c}{Qwen3-Omni} & \multicolumn{2}{c}{Gemini 3} \\
\cmidrule(lr){2-3} \cmidrule(lr){4-5} \cmidrule(lr){6-7} \cmidrule(lr){8-9} \cmidrule(lr){10-11} \cmidrule(lr){12-13}
& Correct & Wrong & Correct & Wrong & Correct & Wrong & Correct & Wrong & Correct & Wrong & Correct & Wrong \\
\midrule
Precision & 0.201 & 0.124 & 0.333 & 0.111 & 0.179 & 0.124 & 0.184 & 0.121 & 0.258 & 0.087 & 0.472 & 0.110 \\
Recall    & 0.188 & 0.116 & 0.327 & 0.109 & 0.264 & 0.183 & 0.419 & 0.275 & 0.369 & 0.124 & 0.478 & 0.111 \\
F1-Score  & 0.195 & 0.120 & 0.330 & 0.110 & 0.213 & 0.148 & 0.256 & 0.168 & 0.304 & 0.102 & 0.475 & 0.110 \\
\bottomrule
\end{tabular}%
}
\end{table*}

\begin{table*}[t]
\centering
\caption{Regional style classification on OverClocked ReMix. Prompt: ``What regional
style would you say this music belongs to?''}
\label{tab:regional}
\begin{tabular}{l cc cc cc cc cc}
\toprule
& \multicolumn{2}{c}{Qwen2-Audio} & \multicolumn{2}{c}{Qwen3-Omni} & \multicolumn{2}{c}{Audio-Flamingo 3} & \multicolumn{2}{c}{Music-Flamingo} & \multicolumn{2}{c}{Gemini 3} \\
\cmidrule(lr){2-3} \cmidrule(lr){4-5} \cmidrule(lr){6-7} \cmidrule(lr){8-9} \cmidrule(lr){10-11}
& Correct & Wrong & Correct & Wrong & Correct & Wrong & Correct & Wrong & Correct & Wrong \\
\midrule
Precision & 0.267 & 0.095 & 0.288 & 0.101 & 0.188 & 0.104 & 0.247 & 0.088 & 0.185 & 0.093 \\
Recall    & 0.099 & 0.035 & 0.262 & 0.092 & 0.110 & 0.061 & 0.206 & 0.073 & 0.188 & 0.095 \\
F1-Score  & 0.145 & 0.052 & 0.275 & 0.096 & 0.139 & 0.077 & 0.224 & 0.080 & 0.186 & 0.094 \\
\bottomrule
\end{tabular}
\end{table*}

\begin{table*}[t]
\centering
\caption{Mood classification on OverClocked ReMix. Prompt: ``What is the mood of this
piece of music?''}
\label{tab:mood}
\begin{tabular}{l cc cc cc cc cc}
\toprule
& \multicolumn{2}{c}{Qwen2-Audio} & \multicolumn{2}{c}{Qwen3-Omni} & \multicolumn{2}{c}{Audio-Flamingo 3} & \multicolumn{2}{c}{Music-Flamingo} & \multicolumn{2}{c}{Gemini 3} \\
\cmidrule(lr){2-3} \cmidrule(lr){4-5} \cmidrule(lr){6-7} \cmidrule(lr){8-9} \cmidrule(lr){10-11}
& Correct & Wrong & Correct & Wrong & Correct & Wrong & Correct & Wrong & Correct & Wrong \\
\midrule
Precision & 0.304 & 0.155 & 0.247 & 0.115 & 0.356 & 0.158 & 0.284 & 0.131 & 0.260 & 0.146 \\
Recall    & 0.263 & 0.133 & 0.487 & 0.226 & 0.352 & 0.156 & 0.324 & 0.150 & 0.403 & 0.226 \\
F1-Score  & 0.282 & 0.143 & 0.328 & 0.152 & 0.354 & 0.157 & 0.303 & 0.140 & 0.316 & 0.177 \\
\bottomrule
\end{tabular}
\end{table*}

\begin{table*}[t]
\centering
\caption{Time signature classification on OverClocked ReMix. Prompt: ``What is the
time signature of this piece of music?''}
\label{tab:timesig}
\begin{tabular}{l cc cc cc cc cc}
\toprule
& \multicolumn{2}{c}{Qwen2-Audio} & \multicolumn{2}{c}{Qwen3-Omni} & \multicolumn{2}{c}{Audio-Flamingo 3} & \multicolumn{2}{c}{Music-Flamingo} & \multicolumn{2}{c}{Gemini 3} \\
\cmidrule(lr){2-3} \cmidrule(lr){4-5} \cmidrule(lr){6-7} \cmidrule(lr){8-9} \cmidrule(lr){10-11}
& Correct & Wrong & Correct & Wrong & Correct & Wrong & Correct & Wrong & Correct & Wrong \\
\midrule
Precision & 0.716 & 0.718 & 0.888 & 0.867 & 0.879 & 0.858 & 0.872 & 0.885 & 0.714 & 0.652 \\
Recall    & 0.621 & 0.622 & 0.866 & 0.846 & 0.858 & 0.858 & 0.582 & 0.591 & 0.728 & 0.666 \\
F1-Score  & 0.665 & 0.667 & 0.877 & 0.856 & 0.869 & 0.869 & 0.698 & 0.709 & 0.721 & 0.659 \\
\bottomrule
\end{tabular}
\end{table*}

A striking feature of the time-signature results is that, for every model
tested, the Correct and Wrong columns are statistically indistinguishable.
This is the cleanest negative finding in our benchmark: even though several
models report time signatures with apparent confidence and high
self-consistency (note the high precision and recall values relative to
chance), their predictions are no more accurate when paired with the actual
audio than when paired with an unrelated track. We interpret this as
evidence that current LALMs cannot reliably perform meter inference from
audio, and instead default to the most common time signature in their
training distribution.

\section{Prompting Sensitivity Studies}
\label{prompting_studies}

The factuality protocol described in the main paper relies on a single
prompt per task. A natural concern is whether the reported numbers are
artifacts of prompt phrasing: a model might appear to fail on a task simply
because it dislikes the way the question is asked. To test this, we ran a
series of supplementary experiments varying the prompt format. These
experiments were conducted on the four LALMs that motivated this work
(LTU-AS, MU-LLaMA, LLaMA-Adapter, SALMONN); we did not extend them to the
newer models because, as the main results show, the newer models'
substantially stronger LLM backbones make them largely insensitive to the
kinds of phrasing variation we test here. We therefore present these
results as evidence about a property of the older generation of LALMs
rather than as a property of the benchmark itself.

\subsection{Linguistic Variation in Open Prompts}

We tested several phrasings of each task's question, ranging from casual
to formal, to measure how much of the older models' performance is a
function of prompt wording rather than audio understanding. We ran four
phrasings each on instrument recognition, genre classification, and
composer classification
(Tables~\ref{tab:musicnet_instrumentation_all_prompt}--\ref{tab:musicnet_composer_all_prompt}).
The first three phrasings of each task are open prompts; the fourth,
marked \emph{(list)}, enumerates the candidate labels in the question.
Each prompt block reports Precision, Recall, and F1-Score under both
Correct (C) and Random (R) audio conditions.

\begin{table*}[t]
\centering
\caption{Linguistic variation on the instrument-recognition prompt
(MusicNet). Each prompt block reports P, R, and F1 under Correct (C) and
Random (R) audio conditions. Prompt~4 is a list prompt that enumerates
the candidate instruments; the others are open prompts. Full prompt text
is given in the footnote.}
\label{tab:musicnet_instrumentation_all_prompt}
\begin{tabular}{@{}llcccccc@{}}
\toprule
& & \multicolumn{2}{c}{LTU-AS} & \multicolumn{2}{c}{MU-LLaMA} & \multicolumn{2}{c}{LLaMA-Adapter} \\
\cmidrule(lr){3-4}\cmidrule(lr){5-6}\cmidrule(lr){7-8}
Prompt & Metric & C & R & C & R & C & R \\
\midrule
\textbf{P1}
 & P  & 0.534 & 0.364 & 0.367 & 0.266 & 0.488 & 0.350 \\
 & R  & 0.461 & 0.314 & 0.582 & 0.422 & 0.676 & 0.486 \\
 & F1 & 0.495 & 0.337 & 0.450 & 0.326 & 0.567 & 0.407 \\
\addlinespace[3pt]
\textbf{P2}
 & P  & 0.502 & 0.316 & 0.302 & 0.246 & 0.458 & 0.325 \\
 & R  & 0.518 & 0.327 & 0.584 & 0.474 & 0.733 & 0.520 \\
 & F1 & 0.510 & 0.321 & 0.398 & 0.324 & 0.564 & 0.400 \\
\addlinespace[3pt]
\textbf{P3}
 & P  & 0.532 & 0.334 & 0.271 & 0.211 & 0.472 & 0.357 \\
 & R  & 0.423 & 0.266 & 0.544 & 0.423 & 0.643 & 0.486 \\
 & F1 & 0.472 & 0.296 & 0.362 & 0.282 & 0.545 & 0.411 \\
\addlinespace[3pt]
\textbf{P4 (list)}
 & P  & 0.155 & 0.149 & 0.199 & 0.172 & 0.164 & 0.164 \\
 & R  & 0.714 & 0.689 & 0.773 & 0.668 & 0.923 & 0.920 \\
 & F1 & 0.254 & 0.245 & 0.316 & 0.273 & 0.279 & 0.278 \\
\bottomrule
\multicolumn{8}{@{}p{0.95\textwidth}@{}}{\footnotesize
\textbf{P1}: ``Which instruments are used in this piece of music?''
\quad
\textbf{P2}: ``Which instruments constitute the instrumentation of this piece?''
\quad
\textbf{P3}: ``What instruments are playing in this song?''
\quad
\textbf{P4}: ``Among \{all instruments\}, which instruments are used in this piece of music?''} \\
\end{tabular}
\end{table*}

\begin{table*}[t]
\centering
\caption{Linguistic variation on the genre-classification prompt (FMA,
same-tree evaluation). Prompt~4 is a list prompt; the others are open
prompts. Full prompt text is given in the footnote.}
\label{tab:fma_genre_all_prompt}
\begin{tabular}{@{}llcccccc@{}}
\toprule
& & \multicolumn{2}{c}{MU-LLaMA} & \multicolumn{2}{c}{LLaMA-Adapter} & \multicolumn{2}{c}{SALMONN} \\
\cmidrule(lr){3-4}\cmidrule(lr){5-6}\cmidrule(lr){7-8}
Prompt & Metric & C & R & C & R & C & R \\
\midrule
\textbf{P1}
 & P  & 0.269 & 0.098 & 0.331 & 0.095 & 0.467 & 0.109 \\
 & R  & 0.306 & 0.112 & 0.339 & 0.097 & 0.220 & 0.051 \\
 & F1 & 0.286 & 0.105 & 0.335 & 0.096 & 0.299 & 0.069 \\
\addlinespace[3pt]
\textbf{P2}
 & P  & 0.218 & 0.086 & 0.272 & 0.077 & 0.256 & 0.083 \\
 & R  & 0.364 & 0.143 & 0.407 & 0.115 & 0.389 & 0.126 \\
 & F1 & 0.273 & 0.107 & 0.326 & 0.092 & 0.309 & 0.100 \\
\addlinespace[3pt]
\textbf{P3}
 & P  & 0.256 & 0.102 & 0.334 & 0.081 & 0.293 & 0.084 \\
 & R  & 0.291 & 0.115 & 0.342 & 0.083 & 0.388 & 0.111 \\
 & F1 & 0.272 & 0.108 & 0.338 & 0.082 & 0.334 & 0.096 \\
\addlinespace[3pt]
\textbf{P4 (list)}
 & P  & 0.201 & 0.124 & 0.333 & 0.111 & 0.179 & 0.124 \\
 & R  & 0.188 & 0.116 & 0.327 & 0.109 & 0.264 & 0.183 \\
 & F1 & 0.195 & 0.120 & 0.330 & 0.110 & 0.213 & 0.148 \\
\bottomrule
\multicolumn{8}{@{}p{0.95\textwidth}@{}}{\footnotesize
\textbf{P1}: ``What is the genre of this song''
\quad
\textbf{P2}: ``What can you infer about the genre of the music''
\quad
\textbf{P3}: ``What genre does this piece of music fall under?''
\quad
\textbf{P4}: ``Among \{all genres\}, what is the genre of this song?''} \\
\end{tabular}
\end{table*}

\begin{table*}[t]
\centering
\caption{Linguistic variation on the composer-classification prompt
(MusicNet). Prompt~4 is a list prompt; the others are open prompts.
Full prompt text is given in the footnote.}
\label{tab:musicnet_composer_all_prompt}
\begin{tabular}{@{}llcccccc@{}}
\toprule
& & \multicolumn{2}{c}{LTU-AS} & \multicolumn{2}{c}{MU-LLaMA} & \multicolumn{2}{c}{LLaMA-Adapter} \\
\cmidrule(lr){3-4}\cmidrule(lr){5-6}\cmidrule(lr){7-8}
Prompt & Metric & C & R & C & R & C & R \\
\midrule
\textbf{P1}
 & P  & 0.232 & 0.238 & 0.181 & 0.179 & 0.268 & 0.240 \\
 & R  & 0.239 & 0.245 & 0.424 & 0.418 & 0.497 & 0.445 \\
 & F1 & 0.236 & 0.242 & 0.254 & 0.250 & 0.348 & 0.312 \\
\addlinespace[3pt]
\textbf{P2}
 & P  & 0.297 & 0.334 & 0.156 & 0.157 & 0.316 & 0.279 \\
 & R  & 0.261 & 0.294 & 0.348 & 0.352 & 0.521 & 0.461 \\
 & F1 & 0.277 & 0.313 & 0.215 & 0.217 & 0.394 & 0.348 \\
\addlinespace[3pt]
\textbf{P3}
 & P  & 0.333 & 0.337 & 0.226 & 0.235 & 0.281 & 0.241 \\
 & R  & 0.285 & 0.288 & 0.618 & 0.642 & 0.506 & 0.433 \\
 & F1 & 0.307 & 0.310 & 0.331 & 0.344 & 0.361 & 0.310 \\
\addlinespace[3pt]
\textbf{P4 (list)}
 & P  & 0.152 & 0.133 & 0.153 & 0.140 & 0.152 & 0.137 \\
 & R  & 0.558 & 0.488 & 0.294 & 0.270 & 0.561 & 0.506 \\
 & F1 & 0.238 & 0.209 & 0.201 & 0.185 & 0.239 & 0.216 \\
\bottomrule
\multicolumn{8}{@{}p{0.95\textwidth}@{}}{\footnotesize
\textbf{P1}: ``Which classical composer's style does this piece resemble the most?''
\quad
\textbf{P2}: ``Which classical composer's compositional style does this piece most closely resemble?''
\quad
\textbf{P3}: ``Which classical composer's work does this sound like?''
\quad
\textbf{P4}: ``Among \{all composers\}, whose style does this piece of music sound like?''} \\
\end{tabular}
\end{table*}

Two patterns emerge consistently across the three tasks. First, the F1
score of each older model drifts by up to roughly 0.10 absolute as the
question wording is varied, even though the underlying task and the
audio are identical---a level of brittleness that complicates any
attempt to summarize a single ``true'' performance number for these
models. Second, the list-prompt variant (the fourth row block of each
table) collapses precision while leaving recall close to 1.0: the older
models respond to an enumerated candidate set by emitting most of the
listed labels rather than discriminating among them. This pattern is
most extreme for instrumentation, where recall exceeds 0.9 against
precision near 0.16 for LLaMA-Adapter. Composer is essentially at chance
for all three models across all phrasings, which is unsurprising given
that none of these models were trained on classical repertoire. We
interpret these results as evidence about a property of the older
generation of LALMs: their narrow instruction-tuning distributions make
them sensitive to surface phrasing in ways that newer models with
substantially stronger LLM backbones are not.

\subsection{Same-Node Genre Classification on FMA}
\label{same_node_fma}

The genre results in Section~\ref{full_factuality_results} use the
\emph{same-tree} criterion, in which a prediction is counted correct if it
matches any node in the same subtree as the ground-truth label. We also
ran a stricter \emph{same-node} variant, in which only an exact match with
the ground-truth label is counted as correct. Results are reported in
Table~\ref{tab:fma_same_node}.

\begin{table*}[t]
\centering
\caption{Same-node genre classification on FMA across three open prompts.}
\label{tab:fma_same_node}
\begin{tabular}{lcccccc}
\toprule
\multicolumn{7}{l}{\textbf{Prompt: ``What is the genre of this song''}} \\
\addlinespace[2pt]
 & \multicolumn{2}{c}{MU-LLaMA} & \multicolumn{2}{c}{LLaMA-Adapter} & \multicolumn{2}{c}{SALMONN} \\
 & Correct & Random & Correct & Random & Correct & Random \\
\cmidrule(lr){2-3}\cmidrule(lr){4-5}\cmidrule(lr){6-7}
Precision & 0.093 & 0.035 & 0.181 & 0.041 & 0.173 & 0.039 \\
Recall & 0.117 & 0.043 & 0.206 & 0.047 & 0.086 & 0.019 \\
F1-Score & 0.104 & 0.039 & 0.193 & 0.044 & 0.115 & 0.026 \\
\addlinespace[6pt]

\multicolumn{7}{l}{\textbf{Prompt: ``What can you infer about the genre of the music''}} \\
\addlinespace[2pt]
 & \multicolumn{2}{c}{MU-LLaMA} & \multicolumn{2}{c}{LLaMA-Adapter} & \multicolumn{2}{c}{SALMONN} \\
 & Correct & Random & Correct & Random & Correct & Random \\
\cmidrule(lr){2-3}\cmidrule(lr){4-5}\cmidrule(lr){6-7}
Precision & 0.034 & 0.034 & 0.120 & 0.026 & 0.101 & 0.022 \\
Recall & 0.157 & 0.064 & 0.249 & 0.055 & 0.244 & 0.054 \\
F1-Score & 0.108 & 0.044 & 0.162 & 0.036 & 0.143 & 0.031 \\
\addlinespace[6pt]

\multicolumn{7}{l}{\textbf{Prompt: ``What genre does this piece of music fall under?''}} \\
\addlinespace[2pt]
 & \multicolumn{2}{c}{MU-LLaMA} & \multicolumn{2}{c}{LLaMA-Adapter} & \multicolumn{2}{c}{SALMONN} \\
 & Correct & Random & Correct & Random & Correct & Random \\
\cmidrule(lr){2-3}\cmidrule(lr){4-5}\cmidrule(lr){6-7}
Precision & 0.084 & 0.033 & 0.184 & 0.038 & 0.122 & 0.027 \\
Recall & 0.128 & 0.050 & 0.217 & 0.045 & 0.231 & 0.052 \\
F1-Score & 0.102 & 0.039 & 0.199 & 0.041 & 0.160 & 0.036 \\
\bottomrule
\end{tabular}
\end{table*}

\subsection{Binary-Choice Format}

To further simplify evaluation in a human-interpretable way, we designed a
binary-choice format in which the model is asked to choose between the
exact ground-truth answer and one randomly selected distractor, framed as:
``Between A and B, \ldots''. To mitigate positional bias, the order was
randomized so that the correct answer appeared first in 50\% of the
prompts and second in the remaining 50\%. The result of binary-choice
experiments on MusicNet and FMA is shown in
Tables~\ref{tab:binary_musicnet} and \ref{tab:binary_fma}.

\begin{table*}[t]
\centering
\caption{Binary-choice composer classification on MusicNet.}
\label{tab:binary_musicnet}
\begin{tabular}{lccc}
\toprule
\multicolumn{4}{l}{\textbf{Prompt: ``Between \{two composers\}, whose style does this piece resemble the most?''}} \\
\addlinespace[2pt]
Metric & LTU-AS & MU-LLaMA & LLaMA-Adapter \\
\midrule
Precision & 0.496 & 0.570 & 0.463 \\
Recall    & 0.709 & 0.655 & 0.803 \\
F1-Score  & 0.584 & 0.609 & 0.588 \\
\bottomrule
\end{tabular}
\end{table*}

\begin{table*}[t]
\centering
\caption{Binary-choice genre classification on FMA.}
\label{tab:binary_fma}
\begin{tabular}{lccc}
\toprule
\multicolumn{4}{l}{\textbf{Prompt: ``Between \{two genres\}, what is a better description of the genre of this piece?''}} \\
\addlinespace[2pt]
Metric & MU-LLaMA & LLaMA-Adapter & SALMONN \\
\midrule
Precision & 0.464 & 0.500 & 0.460 \\
Recall    & 0.740 & 0.590 & 0.853 \\
F1-Score  & 0.570 & 0.542 & 0.598 \\
\bottomrule
\end{tabular}
\end{table*}

The binary-choice scores hover near 0.55--0.60 F1, only modestly above the
0.5 chance baseline. This is an important sanity check: even when the
question is reduced to a forced choice between two options, the older
LALMs perform only slightly better than coin-flipping.

\subsection{True/False Format}

For recognition tasks, we also asked the model to make a true--false
judgment for each possible label in the dataset, producing a table of
boolean values indicating whether the model believed each label was
present in the given audio. From these predictions we computed accuracy,
which accounts for both true positives and true negatives. For chunked
models (LTU-AS, MU-LLaMA, LLaMA-Adapter, which process audio in
60-second segments), we report two aggregation strategies:

\begin{enumerate}
    \item \textbf{Inclusive (any-chunk):} the final response is true if
    the model returns true on \emph{any} chunk. This strategy ensures that
    instruments appearing in only a short section of the music are still
    detectable. Results are shown in
    Table~\ref{tab:ture_false_musicnet_include}.

    \item \textbf{Majority:} the final response is true if the model
    returns true on a \emph{majority} of chunks. This strategy reveals
    whether a model has a stable bias toward one answer rather than
    random guessing. Results are shown in
    Table~\ref{tab:ture_false_musicnet_majority}.
\end{enumerate}

\begin{table*}[t]
\centering
\caption{True/false instrumentation on MusicNet (any-chunk aggregation).}
\label{tab:ture_false_musicnet_include}
\begin{tabular}{lcccccc}
\toprule
\multicolumn{7}{l}{\textbf{Prompt: ``Is \{instrument\} used in this song?''}} \\
\addlinespace[2pt]
Metric & \multicolumn{2}{c}{LTU-AS} & \multicolumn{2}{c}{MU-LLaMA} & \multicolumn{2}{c}{LLaMA-Adapter} \\
& Correct & Wrong & Correct & Wrong & Correct & Wrong \\
\cmidrule(lr){2-3}\cmidrule(lr){4-5}\cmidrule(lr){6-7}
Precision & 0.182 & 0.171 & 0.179 & 0.171 & 0.179 & 0.174 \\
Recall    & 0.670 & 0.628 & 0.876 & 0.839 & 0.876 & 0.854 \\
F1-Score  & 0.287 & 0.268 & 0.297 & 0.284 & 0.297 & 0.289 \\
Accuracy  & 0.167 & 0.149 & 0.168 & 0.167 & 0.174 & 0.166 \\
\bottomrule
\end{tabular}
\end{table*}

\begin{table*}[t]
\centering
\caption{True/false instrumentation on MusicNet (majority aggregation).}
\label{tab:ture_false_musicnet_majority}
\begin{tabular}{lcccccc}
\toprule
\multicolumn{7}{l}{\textbf{Prompt: ``Is \{instrument\} used in this song?''}} \\
\addlinespace[2pt]
Metric & \multicolumn{2}{c}{LTU-AS} & \multicolumn{2}{c}{MU-LLaMA} & \multicolumn{2}{c}{LLaMA-Adapter} \\
& Correct & Wrong & Correct & Wrong & Correct & Wrong \\
\cmidrule(lr){2-3}\cmidrule(lr){4-5}\cmidrule(lr){6-7}
Precision & 0.181 & 0.166 & 0.190 & 0.191 & 0.192 & 0.189 \\
Recall    & 0.659 & 0.607 & 0.601 & 0.605 & 0.795 & 0.780 \\
F1-Score  & 0.284 & 0.261 & 0.288 & 0.290 & 0.310 & 0.304 \\
Accuracy  & 0.165 & 0.152 & 0.168 & 0.164 & 0.183 & 0.176 \\
\bottomrule
\end{tabular}
\end{table*}

The true/false format collapses precision sharply across all three older
models: when asked label-by-label about every instrument in the
vocabulary, the models tend to say ``true'' for nearly all of them,
yielding high recall but precision near the base rate. The Correct--Wrong
gap is also smaller than under the open prompt, indicating that
instruction-following ability, rather than audio understanding, is the
limiting factor in this format for older LALMs. Taken together with the
linguistic-variation and binary-choice studies above, these results
support our use of the open and list prompts as the primary evaluation
formats in the main paper.

\end{document}